\documentclass[trackchanges]{aastex701}

\shorttitle{Digital Calibration Source for 21\,cm Array}
\shortauthors{Bhopi et al.}

\received{XXX}
\revised{YYY}
\accepted{ZZZ}

\usepackage{graphicx}
\usepackage{amsmath}

\newcommand{\tcm}{21\,cm~}
\newcommand{\son}{\textbf{ON} }
\newcommand{\soff}{\textbf{OFF} }

\newcommand{\tpeacc}{$T_{\mathrm{PEACC}}$ }
\newcommand{\tsys}{$T_{\mathrm{sys}}$ }

\newcommand{\peacc}{PEACC }
\newcommand{\m}[1]{\mathrm{#1}}

\graphicspath{{./}{figures/}}

\begin{document}

\title{PEACC --- Precision Emitter for 21\,cm Array Coherent Calibration}

\author[orcid=0000-0002-9218-1624,gname=Kalyani,sname=Bhopi]{Kalyani Bhopi}
\email[show]{kalyani.bhopi@mail.wvu.edu}
\correspondingauthor{Kalyani Bhopi}
\affiliation{Lane Department of Computer Science and Electrical Engineering,
West Virginia University, Morgantown, WV, USA}
\affiliation{Center for Gravitational Waves and Cosmology,
West Virginia University, Morgantown, WV, USA}

\author[orcid=0009-0001-9212-5761,gname=Morgan,sname=Cole]{Morgan Cole}
\email{morgan.cole@yale.edu}
\affiliation{Department of Physics, Yale University, New Haven, CT, USA}

\author[orcid=0009-0009-2936-4821,gname=Mallory,sname=Helfenbein]{Mallory Helfenbein}
\email{mallory.helfenbein@yale.edu}
\affiliation{Department of Physics, Yale University, New Haven, CT, USA}
\affiliation{Department of Astronomy, University of Virginia, Charlottesville, VA, USA}

\author[orcid=0000-0001-6196-5955,gname=Will,sname=Tyndall]{Will Tyndall}
\email{will.tyndall@yale.edu}
\affiliation{Department of Physics, Yale University, New Haven, CT, USA}
\affiliation{Trottier Space Institute, McGill University, Montreal, QC, Canada}

\author[orcid=0000-0000-0000-0000,gname=Audrey,sname=Whitmer]{Audrey Whitmer}
\email{audrey.whitmer@yale.edu}
\affiliation{Department of Physics, Yale University, New Haven, CT, USA}

\author[orcid=0000-0003-3772-2798,gname=Kevin,sname=Bandura]{Kevin Bandura}
\email{kevin.bandura@mail.wvu.edu}
\affiliation{Lane Department of Computer Science and Electrical Engineering,
West Virginia University, Morgantown, WV, USA}
\affiliation{Center for Gravitational Waves and Cosmology,
West Virginia University, Morgantown, WV, USA}

\author[orcid=0000-0002-7333-5552,gname=Laura,sname=Newburgh]{Laura Newburgh}
\email{laura.newburgh@yale.edu}
\affiliation{Department of Physics, Yale University, New Haven, CT, USA}

\begin{abstract}
Foreground mitigation remains a central challenge for \tcm intensity mapping experiments, which require precise, wideband calibration of telescope beams and gains. We present the Precision Emitter for \tcm Array Coherent Calibration (PEACC), a digitally synthesized calibration source that generates Gaussian noise across a 1.2\,GHz bandwidth, time-synchronized to a 1\,pulse-per-second output from a GPS-disciplined oscillator, and optimized for aerial deployment. \peacc uses a dual-source architecture with one unit mounted on an aerial platform and a second reference unit connected directly to the radio data acquisition system; this configuration enables improved sensitivity in the low-SNR regime and direct phase measurement. The system further supports configurable band selection, allowing adaptation to various \tcm intensity mapping telescopes. We validated \peacc through anechoic chamber measurements and by integrating the source on a drone flown over a local radio dish testbed. In both settings, the correlated channel substantially outperformed the auto-correlation channel across all signal-to-noise regimes of interest, confirming the key advantage of the dual-source architecture. To our knowledge, this is the first published demonstration of a free-space coherent calibration signal synchronized only by clocks, the first deployment of such a source on a drone, and the first published beam measurements made with such a source. Given the growing interest in drone-based calibration for \tcm arrays, this work establishes the feasibility of high-fidelity digital calibration for next-generation \tcm instruments, and provides a practical path towards improved foreground control and beam calibration in future arrays.
\end{abstract}



\section{Introduction}
\label{sec:intro}

\tcm radio surveys have emerged as a promising tool over the past few decades to measure cosmological neutral hydrogen across a variety of redshifts with a broad range of science goals and experimental
configurations (HERA~\citep{DeBoer:2016tnn,Berkhout2024HERAPhaseII}, CHIME~\citep{Bandura:2014, CHIMEoverview}, HIRAX~\citep{Newburgh2016_HIRAX_SPIE, 2022JATIS...8a1019C}, SARAS3~\citep{2021arXiv210401756N}, REACH~\citep{Roque2025REACH}, EDGES~\citep{2017ApJ...835...49M}, PRIZM~\citep{2019JAI.....850004P}, SKA~\citep{VillaescusaNavarro2017SKABAO}, LOFAR~\citep{vanHaarlem2013}, MWA~\citep{Tingay2013MWA}, PAPER~\citep{Parsons2010PAPER, Parsons2014PAPER32}, DSA~\citep{Hallinan2019DSA2000, Mahesh2023DSA200021cm}, BMX~\citep{OConnor2020BMX}, BINGO~\citep{Wuensche2021BINGOProjectI, Wuensche2021BINGOStatus}, and upcoming CHORD~\citep{Vanderlinde:2019tjt}). Separating the \tcm signal from bright synchrotron foreground emission is a crucial step toward measuring cosmological signals~\citep{Kerrigan:2018rfz, Kohn:2016xwd, Thyagarajan:2015ewa}, but in practice the chromatic response of the telescope impedes this separation, requiring precision characterization of the frequency-dependent beam shape~\citep{Shaw:2014vy, Thyagarajan:2015ewa, LiuShaw2020Review}. 

Compact \tcm interferometers designed to measure the faint ($\sim$100\,mK) diffuse signals lack the spatial resolution and scanning to employ traditional point-source based beam calibration. As a result, many \tcm stationary drift scanning experiments are developing instrumentation, especially radio sources carried by drones, to directly measure the instrumental beam shape~\citep{Tianlai_drones, Chang_drones, Virone_drones, Jacobs_drones, Pupillo_drones, Newburgh:2014, Kuhn_2025, Tyndall_2025, Tianlai_2026, BRAMS}.

In our previous work, we described a digital calibration source, ideal for drone deployment, which would generate a coherent, correlate-able free-space signal~\citep{Bhopi_2022}. Conceptually, this calibrator is a pair of sources: each uses a pulse-per-second trigger to generate an identical copy of the signal from a pre-determined seed. The first source is flown on the drone (`transmission') and measured through the telescope by the instrument correlator, the second source is connected directly to the instrument correlator (‘reference’). Because the reference signal is a known, deterministic copy of the transmitted signal, the cross-correlation of the reference against each telescope input isolates the calibration signal directly. These measurements can be made without requiring background subtraction. This enables simultaneous measurement of both amplitude and phase without the noise penalty incurred by the background subtraction that switched or pulsed incoherent sources~\citep{2017ExA....43..119P} require. Simulations in~\citet{Bhopi_2022} demonstrated that this approach yields a higher signal-to-noise ratio than such switched sources, with the greatest improvement in the faint sidelobe regions of the beam. The primary limitation identified was relative timing jitter between the transmission and reference boards, which can slightly decorrelate the calibration signals and hence reduce signal to noise. As a result, our results presented in this paper include measurements of timing jitter, both directly and inferred from beam measurements, which show we meet or exceed the 1.7\,ns requirement found in that paper, equivalent to when the cross-correlation channel meets a 1\% requirement~\citep{Shaw:2014vy} at lower beam amplitudes than the autocorrelation.

In this paper, we present the design, deployment, and testing of \peacc --- a digital noise source on a Xilinx RFSoC platform, integrated with an off-the-shelf timing system, and validate its performance across multiple testing environments ---- including anechoic chamber trials and a local radio dish testbed. Our digital calibration source is designed to operate over a wide bandwidth of up to 1.2\,GHz, covering observing bands of CHIME and HIRAX (400\,MHz-800\,MHz) as well as CHORD (300\,MHz-1.5\,GHz), and supports configurable band selection, allowing adaptation to various \tcm intensity mapping telescopes. We adopted Gaussian noise as a calibration signal to avoid unintended interference with ongoing observations and to ensure the source remains undetectable by other radio instruments. Using a combination of benchtop measurements, beam measurements of an antenna in an anechoic chamber, and drone measurements of a 3\,m testbed radio dish, we demonstrated we meet the 1.7\,ns jitter requirement such that the correlated measurement remains superior to that obtained from a pure noise source alone,  
and estimate we make measurements with statistical precision of 1\% or better down to signal-to-noise level of $\sim$-9\,dB,  
limited only by the dynamic range of the instrument correlator.


The paper is organized as follows. Section~\ref{sec:NSoverview} details the development and implementation of the calibration module on a Xilinx Ultrascale++ RFSoC platform\footnote{\href{https://www.amd.com/en/corporate/university-program/aup-boards/rfsoc4x2.html}{Xilinx RFSoC4x2}} including characterization of the GPS-disciplined oscillator used for the timing synchronization, and laboratory measurements of clock stability and drift. 
Section~\ref{sec:benchtop_valid} presents initial bench test results, Section~\ref{sec:chamber_measure} describes beam measurements conducted in a controlled anechoic chamber, and Section~\ref{sec:drone_demo} describes the drone, its payload, drone flights performed during beam measurements, and beam pattern measurements of a 3\,m radio test dish. Lastly, Section~\ref{sec:outlook} summarizes the results from this paper, highlighting the successful achievement of beam amplitude recovery at low signal-to-noise, placing an all-sky 1\% beam measurement firmly within reach. 


\begin{figure*}[ht!]
\centering
\includegraphics[width=0.8\textwidth, height=0.3\textwidth]{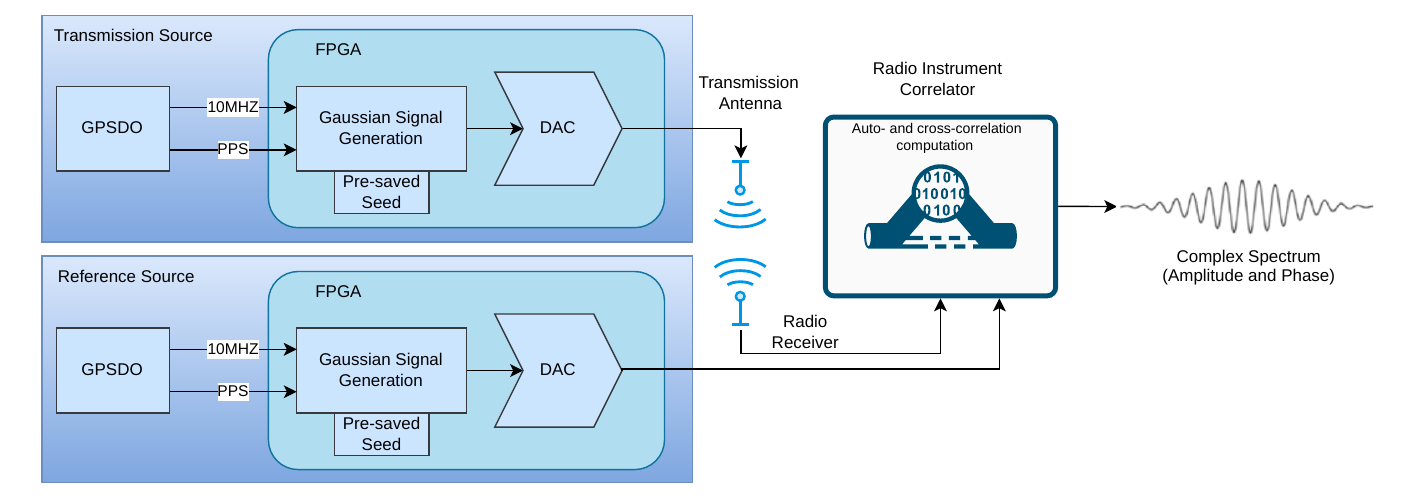}
\caption{\small{\textit{Illustration of the digital calibration source concept, described in detail in the text. In each of the two calibration boards, a GHz-wide Gaussian complex signal is generated once per second in the FPGA and converted to analog using the on-board DAC, with generation triggered by an external PPS and seeded from FPGA memory. One signal is transmitted into the telescope aperture, the other signal is directly attached to the correlator as a reference. The resulting complex signal can be detected with high signal-to-noise after correlation in the instrument correlator.}}}
\label{fig:dns_schematic}
\end{figure*}

\section{Digital Calibration Source}
\label{sec:NSoverview}

The conceptual design for the free-space calibration module was described in \citet{Bhopi_2022} and shown in Figure~\ref{fig:dns_schematic}. Briefly, the calibration system employs two identical calibration source units. One copy is transmitted from the drone-mounted module into the radio dish aperture, while an identical copy is locally generated and directly attached as an analog input to the correlator of the radio instrument. Each calibration source incorporates a Field Programmable Gate Array (FPGA) that receives a pulse-per-second (PPS) signal from an external clock synchronized to the GPS satellites. Upon every PPS trigger, the firmware loads a pre-saved deterministic seed from the FPGA memory, ensuring generation of the same pseudo-random sequence each second. This sequence, although statistically random, is completely reproducible as it is defined by the stored seed value. The resulting complex (real and imaginary) digital signal is converted to analog form using an on-board DAC. The DAC configuration can be dynamically adjusted in real time, allowing changes to parameters such as sampling rate and mixer carrier frequency through firmware updates. Since both calibration units generate signals from the same seed and GPS-synchronized timing, the transmitted and locally generated signals are exact replicas. These two signals are subsequently correlated within the radio data acquisition system, yielding the instrument’s complex response to the calibration signal for each antenna element at the same spectral resolution as the instrument. The correlated output remains deterministic as it originates from identical, seed-defined sequences in both FPGA modules. 

In~\citet{Bhopi_2022}, we presented a prototype implementation based on the LimeSDR\footnote{\href{https://limemicro.com/boards/limesdr/}{LimeSDR}} (software defined radio) platform. In this initial work, we demonstrated that the system was capable of achieving a high signal-to-noise ratio (SNR) in the correlated channel. However, this approach faced several notable limitations, including constraints on bandwidth, the necessity for an external, tightly synchronized timing reference to meet strict timing requirements, a significant lag of $\sim 5$ ms between input channels, silent packet drops, and considerable gain drift as a function of temperature. These technical challenges motivated us to pursue a revised implementation, as shown schematically in Figure~\ref{fig:dns_schematic}. The updated \peacc components, described in more detail in this section, consist of a Xilinx Ultrascale+ RFSoC which handles the digital signal generation every second upon reset with a PPS signal, 10\,MHz timing synchronization with its on-board clock registers, and conversion to analog signal with on-board DAC. The PPS reset and 10\,MHz square wave for synchronization are provided by a commercial GPS-disciplined oscillator (GPSDO) --- a device in which an on-board crystal oscillator is continuously phase-locked to the PPS output of a GPS receiver, transferring the long-term frequency stability of GPS atomic clocks to a local, low-cost oscillator. 

\subsection{RFSOC} 
\label{sec:rfsoc}

To satisfy the technical demands of a GHz-broadband \peacc and simplify practical realization of our theoretical framework, we required a more advanced chipset than the LimeSDR. We identified several essential criteria for suitable hardware: 
\begin{itemize}
\item Sufficient bandwidth that allows high sampling rates to support GHz-scale signal generation
\item Extensive FPGA resources to accommodate complex digital signal processing chains
\item Precise synchronization capabilities with external timing references and dedicated input for external pulse-per-second (PPS) synchronization
\item Integrated, high-performance digital-to-analog conversion (DAC)
\item Flexible support for real-time updates such as carrier frequency adjustment, DAC output power control, and PLL tuning
\item An efficient programming environment to streamline the development workflow
\end{itemize}
Both the Xilinx Zynq UltraScale+ RFSoC XCZU28DR-2FFVG1517E, implemented on the RFSoC2x2 board \footnote{\href{https://www.amd.com/en/corporate/university-program/aup-boards/rfsoc2x2.html}{Xilinx RFSoC2x2}}, and the XCZU48DR-2FFVG1517E, used on the RFSoC4x2 board, were evaluated against these criteria and found to be well-suited for our application. Ultimately, the RFSoC4x2 platform was selected as the primary development environment, in part because it was available during the testing period. 

For hardware synthesis, design validation, and simulation, we used AMD-Xilinx's Vivado tool suite, which provides the standard environment for developing programmable logic overlays and managing hardware resources on the RFSoC platform. Furthermore, the PYNQ (Python Productivity for Zynq) framework was utilized to interact with the programmable logic, manage hardware overlays, and configure DAC parameters as well as on-board PLL registers through high-level Python code executed in JupyterLab running on the on-board ARM processor. This approach significantly reduced both design complexity and development time by eliminating conventional hardware-centric design workflows. 

\subsection{RFSoC clock synthesizers and Pulse-Per-Second trigger} 
\label{sec:fpga_clocking}
The \peacc timing architecture relies on two distinct signals supplied by the external GPS-disciplines oscillator (GPSDO): a 10\,MHz reference clock for signal generation and a pulse-per-second (PPS) trigger for periodic synchronization. These are processed separately on-board and together ensure coherence between the transmission and reference boards. 

The 10\,MHz reference undergoes an additional on-board stage of jitter cleaning and signal conditioning via the LMK04828 and LMX2594 clock chips.
This clock distribution path sequentially uses an external GPSDO, followed by the LMK04828 and LMX2594 clock chips, before delivery to the FPGA fabric. Both clock chips are specified to have 91\,fs jitter or better across our signal generation bandwidth~\citep{TI_LMK04828EP_DS, TI_LMX2594_DS}, which is sufficiently low jitter for our synchronization requirements. This ensures the generation of stable and accurate clock signals required by both the FPGA fabric and the RF/sampling subsystems on-board, which is a crucial factor in synchronization of the whole system. 

The PPS signal from the GPSDO is routed through the board's comparator path into the FPGA logic, where it is used to reset the calibration signal once per second. This periodic reset mitigates the effects of long-term clock drift and maintains continuous synchronization with the reference board.

\begin{figure*}[ht!]
\centering
\includegraphics[width=0.6\textwidth, height=0.47\textwidth]{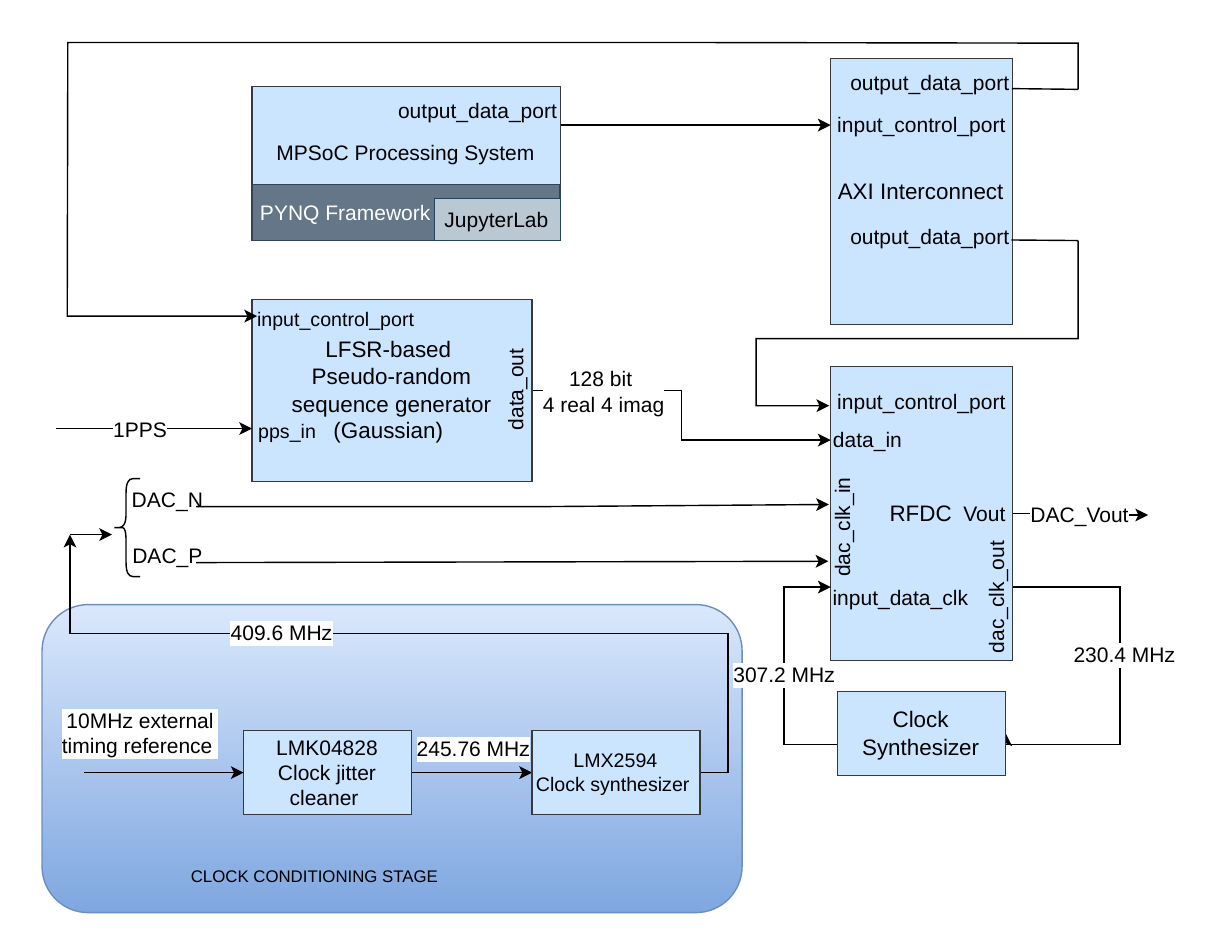}
\includegraphics[width=0.37\textwidth, height=0.47\textwidth]{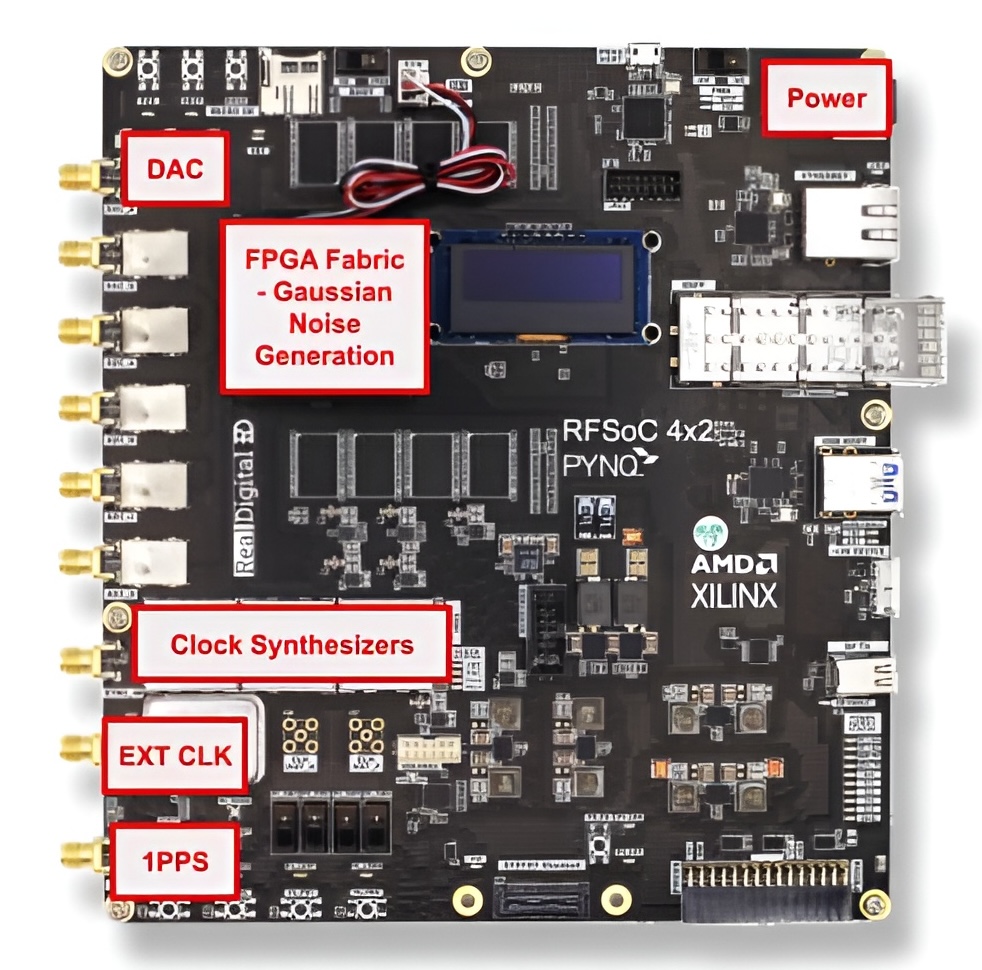}
\caption{\small{\textit{System architecture and corresponding blocks of
the \peacc source on the RFSoC4$\times$2 platform. (a) Block diagram
showing the signal processing chain: the clock conditioning stage
processes the incoming 10\,MHz external timing reference, up-converting
it successively to 245.76\,MHz and 409.6\,MHz to discipline the RFDC
(RF Data Converter) DAC block. The Gaussian noise generator block accepts
1\,PPS input to reset the deterministic signal every second, generating
128-bit wide data comprising four real and four imaginary 16-bit
components, which are then fed to the RFDC block for processing from
complex to real format prior to output at the DAC port. Inter-block
communication occurs via the AXI protocol under control of the MPSoC
processing system (ARM processor on RFSoC4$\times$2), with the PYNQ
framework integrated to enable JupyterLab-based runtime control of the
entire system. (b) Photograph of the RFSoC4$\times$2 evaluation board
\citep{realdigital2024rfsoc}, with key components used in \peacc
development indicated.
}}}
\label{fig:system_architect}
\end{figure*}

\subsection{Development of Gaussian-distributed signal}
\label{sec:gaussian_dev}

Our high-performance Digital Gaussian Noise Generator is designed using  Xilinx System Generator for DSP, a high level design environment integrated with MATLAB/Simulink. At the core of the system is a pseudo-random number generation scheme that sums the outputs of multiple 128-bit linear feedback shift register (LFSR) units~\citep{Golomb1982}. We adopted this approach over box-muller or rejection sampling primarily because our application requires a spectrally flat wideband RF noise source - not a precise Gaussian amplitude distribution. 
Summing multiple de-correlated LFSR-based sequences produces a broadband spectrum with a spectrally flat power spectral density (PSD), since the purely additive operation preserves the flat PSD of the individual LFSR outputs~\citep{Proakis2007}. Non-linear transforms such as Box-Muller, by contrast, introduce spectral coloring artifacts that would degrade the flatness of the output spectrum. 
The shift register and adder structures scale naturally to high clock rates, map efficiently onto FPGA fabric, and produce exactly one output sample per clock cycle, a property that rejection-based methods cannot guarantee~\citep{Knuth1997}, making them unsuitable for our application where a fixed, known number of samples must be produced per pulse. Although the total resource utilization of this approach is broadly comparable to that of Box-Muller or other alternatives, the key difference lies in spectral performance and sustained throughput: the LFSR-sum approach achieves a flat output spectrum and a guaranteed constant sample rate per clock cycle, at the cost of producing an Irwin-Hall amplitude distribution rather than a true Gaussian. 
The Irwin-Hall (central-limit-theorem based) form~\citep{Marengo2017} exhibits bounded tails relative to a true Gaussian. This represents a deliberate and acceptable trade-off, since our application places no requirement on accurate tail behavior. 

The noise generation architecture operates in two stages. In the first stage, 8 random bits from each 128-bit LFSR instance are extracted per cycle. To enhance randomness and approach a Gaussian-like distribution through the central limit theorem, 32 such LFSR instances are summed together. This addition produces a 13-bit intermediate result, from which the lower 8 bits are retained; this constitutes a single noise block output. In the second stage, 8 such blocks are evaluated in parallel --- each drawing on its own set of 32 LFSRs, for a total of 256 LFSR instances across the full design. The eight 8-bit block outputs are summed to produce an 11-bit intermediate value, which is then zero-padded with 5 most significant bits to yield the final 16-bit output sample per clock cycle. This zero-padding does not alter the statistical profile of the noise but ensures alignment with the fixed-point word length required by downstream processing blocks. To achieve high throughput and meet stringent timing constraints, the design leverages Xilinx DSP48 slices to implement all arithmetic operations. The complete system comprises eight independent signal chains, each generating a 16-bit wide sample, resulting in a 128-bit total data width. These samples are logically grouped as four real and four imaginary components, each 16 bits wide, suitable for complex baseband processing. In practice, the resulting amplitude distribution matches a Gaussian with high fidelity in the central $\sim$$1\sigma$ region as quantified by kurtosis = -0.20 (lighter tails) and maximum standard deviation of 0.0244 between the given and best-fit Gaussian CDF over $|x| < 1\sigma$, which is sufficient for our calibration use case, while the extreme tails are intentionally de-emphasized in favor of wideband spectral flatness and implementation simplicity.  

Timing synchronization is governed by the PPS as described in Section~\ref{sec:fpga_clocking}. Each 128-bit LFSR is seeded with a unique 128-bit initialization value stored in memory. On the rising edge of the PPS, the stored seed is loaded, and the LFSR begins generating random sequences for the duration of the high phase of the PPS signal. When the signal returns low, the system halts the sequence progression and holds the seed value at the output until the next rising edge. To ensure deterministic behavior and maintain synchronization across the system, delay elements are inserted along the data pipeline. These serve to relax critical timing paths and align processing stages, further contributing to stable operation at higher frequencies. For system integration, the design incorporates both the AXI4-Stream master and slave interfaces, enabling seamless communication with the broader FPGA-based processing system, which is built around the AXI protocol. 





\begin{figure*}[ht!]
\centering

\begin{minipage}[c]{0.03\textwidth}
    \centering\textbf{(a)}
\end{minipage}%
\begin{minipage}[c]{0.94\textwidth}
    \includegraphics[width=0.95\linewidth, height=0.35\textheight]{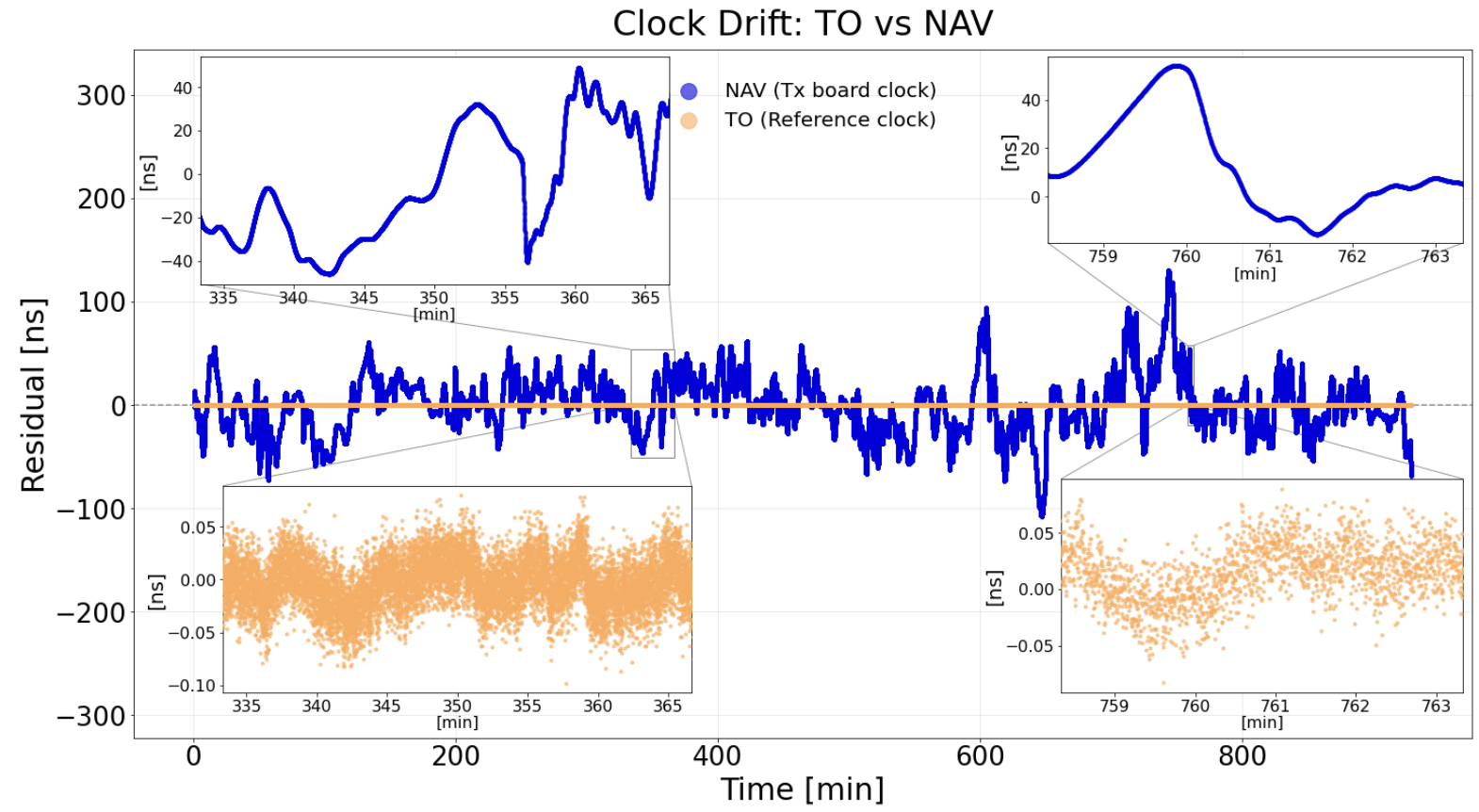}
\end{minipage}

\medskip

\begin{minipage}[c]{0.03\textwidth}
    \centering\textbf{(b)}
\end{minipage}%
\begin{minipage}[c]{0.94\textwidth}
    \includegraphics[width=0.95\linewidth, height=0.35\textheight]{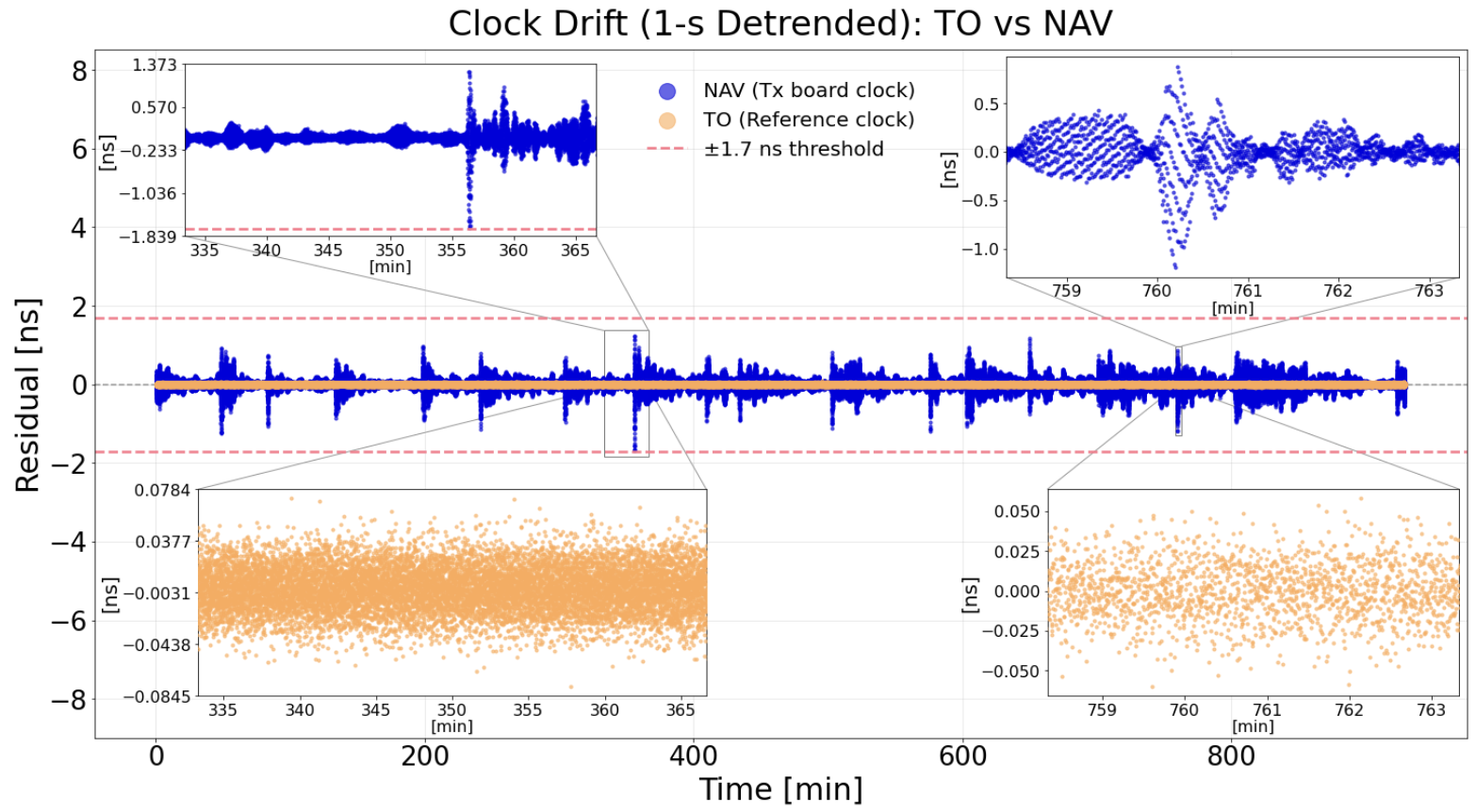}
\end{minipage}

\caption{\small{\textit{Jitter timestream over the approximately 13 hour measurement period for the reference TO mode clock and the transmission NAV mode clock (a) and the per-second detrended timestreams calculated by dividing out the mean value of each 1s interval (b). The x-axis displays the elapsed time in minutes of the data run and the y axis is the jitter value in nanoseconds determined as the residual or offset from 10MHz in a 169.972ms integration period. The plots represent the results from the TO mode clock that is disciplining the correlator and would discipline the stationary Xilinx reference board during calibration measurements (orange). The independent NAV mode clock which would discipline the mobile transmission board during calibration measurements is also overlaid on the same axes (blue). In plot (a) we see the slight effects of large scale drift in plot in the TO mode reference clock and significant drifting behavior by comparison in NAV mode clock, as expected. In plot (b) the detrending mitigates the macroscopic drifting effects, yet obvious signatures of the significant excursions in the NAV timestream in plot (a) can be seen in plot (b), appearing semi-periodic, roughly every 1000 seconds, though upon inspection the signature of each excursion is quite variable. Plot (b) also shows the horizontal lines indicating the 1.7\,ns jitter requirement. It is clear that some rapid drift excursions in the NAV clock data in the transmission board clock produce jitter that approaches but does not exceed this limit.}}}
\label{fig:timingtests}
\end{figure*}

\subsection{Implementation and Integration on RFSoC Architecture}
\label{sec:noise_tx}

Figure~\ref{fig:system_architect} illustrates the overall system architecture of the \peacc source as implemented on the RFSoC4$\times$2 platform. The design integrates clock conditioning, deterministic noise generation, and RF data conversion into a single platform, with the GPSDO providing the external timing reference that synchronizes the system to GPS atomic clocks. The PYNQ framework provides runtime control without hardware recompilation; this enables rapid reconfiguration during field deployment, which is a practical requirement for a drone-based calibration instrument. 
The implementation and deployment of the proposed digital noise generation architecture were carried out using the Xilinx Vivado Design Suite. The noise transmitter intellectual property (IP) core, a reusable, self-contained hardware design module, along with its associated HDL code, was imported into Vivado's custom IP repository. Integration of this core into a broader system-level architecture was achieved using Vivado IP integrator. The system architecture includes key modules such as: Noise generation core, RF Data Converter (RFDC) and Zynq UltraScale+ MPSoC Processing System. These components are interconnected using AXI-based protocols, specifically AXI4-Stream for high-throughput data paths and AXI4-Lite for control and configuration. Communication between modules is managed through AXI interconnect IPs. 

In the proposed architecture, the RFDC (Radio Frequency Data Converter) IP core serves as a critical component for transforming digitally synthesized noise signals into their high-frequency analog counterparts suitable for RF transmission. The 128-bit parallel data stream is sampled at a rate of 3.6864\,GHz, thus defining a Nyquist bandwidth limit of 1.8432\,GHz. This ensures accurate digital-to-analog conversion of the noise signal without aliasing and supports high-fidelity wideband noise transmission. Special emphasis is placed on achieving timing closure for critical timing paths, notably those involving DSP chains and memory interfaces, as these are typically the most sensitive to timing violations due to their high operational frequency and data throughput requirements.

We adapted several existing hardware overlays developed by the StrathSDR group at the University of Strathclyde, which provide basic control functionality via the PYNQ framework~\citep{StrathSDR_RFSOC_Radio_2020}, and integrated them into our design. We incorporated synthesized bitstream and corresponding hardware definition files into a custom PYNQ overlay, which we deployed on an RFSoC platform operating with a Linux-based PYNQ image. Utilizing the Jupyter Notebook environment, we developed interactive dashboards that enable real-time system control and signal monitoring, all without the need for hardware recompilation. This hardware-software co-design methodology significantly expedites the prototyping and validation process for signal processing algorithms, with Python simplifying low-level control to enhance accessibility. 

\subsection{Precision Clocking}
\label{sec:precision_clocking}

As noted above, each external clock synchronizing the two RFSoCs must have a PPS trigger and jitter of 1.7\,ns or better, as established in ~\citet{Bhopi_2022} and discussed in Section~\ref{sec:intro}. Lightweight, low-power GPSDOs offer a practical solution for this requirement. 
We chose to evaluate four different GPSDOs with jitter 
that met our requirements according to their specifications: Brandywine Communications GPSDO Module\footnote{\href{https://www.psirep.com/products/brandywine-gpsdo-gps-disciplined-oscillator-module}{Brandywine GPSDO module}}, Furuno GF-8804 Multi-GNSS Disciplined Oscillator\footnote{\href{https://www.furuno.com/en/products/gnss-module/GF-8805}{Furuno GF-8804}}, Orolia GXClok-500 GPS/GNSS Clock Module\footnote{\href{https://safran-navigation-timing.com/wp-content/uploads/2023/09/GXClok_500_Manual_191222.pdf}{Orolia GXClok-500 GPS}}, and Spectrum Instruments Intelligent Reference/TM-4\footnote{\href{https://www.spectruminstruments.net/products/tm4/tm4.html}{Spectrum Instruments TM-4}}. Beyond jitter performance, several other parameters were also critical, including PPS accuracy, support for multiple GNSS constellations, satellite lock and holdover period, and software configurability. Ultimately, we chose to use the Furuno in lab tests based on its PPS accuracy ($\leq$ 15\,ns), light weight, low power draw, and ease of use. 

Since jitter is the primary determinant of signal quality in this dual-source calibration architecture, we assess the relative jitter of the Furuno clocks in detail. We perform direct measurements of both clocks with an ICEBoard~\citep{Bandura2016} to sample their 10\,MHz outputs. One of the Furuno clocks was set to output two 10\,MHz signals: one square wave which was used to discipline the ICEBoard, and a sine wave which was connected to an analog input port and was sampled directly. The second Furuno clock was set to output only a 10\,MHz sine wave, which was sampled on a second analog input port on the ICEBoard. The default mode of the Furuno GF-8804 clock is “Self-Survey (SS)" mode, which acquires updating timing and position information each second. During drone measurements with this architecture, the Furuno clock disciplining the ICEBoard acts as the stationary reference clock and thus is instead set to “Time Only (TO)” mode which assumes a fixed position for the clock and only acquires updating timing information. The second clock would then be deployed on the drone with the board transmitting the calibration signal and is therefore set to “Navigation (NAV)” mode, which continually acquires information about the position and motion of the clock as well as timing information. We configured the ICEBoard to acquire uncorrelated voltage at the highest possible cadence, acquiring 1 subframe every 20\,ms, each subframe having a 2.56\,$\mu$s integration period. The clocks' sine wave signals were measured for a continuous period of about 13 hours in order to monitor their long term stability. The sine waves are measured via the 8-bit raw ADC datastream of the ICEboard.

The timing jitter of each clock is determined by performing a Fast Fourier Transform (FFT) on each subframe of the 10\,MHz sine wave and calculating the phase offset (\citet{DigitalNoiseSource_WVURAIL}). For each clock, we average the resulting phase jitter over a 169.972\,ms period, matching the analysis conditions used to establish the 1.7\,ns requirement ~\citep{Bhopi_2022}. Typical GPSDO clocks have large drifts in phase over timescales greater than a few seconds. However, due to the 1\,second reset of the calibration signal (described in Section~\ref{sec:fpga_clocking}), we are not sensitive to phase drifts on timescales longer than this. As a result, we subtract the mean of every 1 second interval, which removes the effect of large scale drift, as seen in Figure \ref{fig:timingtests}. 
Figure~\ref{fig:timingtests} shows the results from this timing test for each clock over 13 hours. This measurement is fundamentally timing jitter relative to the clock which disciplines the correlator (in this case, the `TO mode' clock). As a result, the jitter from the TO mode clock is expected to be minimal, which is supported by our measurements: even before correction the largest excursion is 0.15\,ns. The uncorrected NAV mode clock drifts significantly, with excursions on the order of 100\,ns. After mean subtraction, the standard deviation over the entire time period is 0.1\,ns, however we find excursions that occasionally approach the 1.7\,ns limit. This confirms that our clocks meet the performance requirements for cross-correlated beam calibration, but we may see decreased SNR due to occasional periods of larger jitter. Finally, the significant drift of GPSDO clocks on timescales longer than several seconds means that obtaining geometric phase information directly from this system will not likely be possible without further development. Because the geometric phase must first be recovered before the electric-field beam phase can be extracted, this degrades the phase information obtainable from the drone beam maps; we return to this point when reporting the drone flight results in Section \ref{sec:phaseandlims}. 


\begin{figure*}[ht!]
\centering
\includegraphics[width=\textwidth]{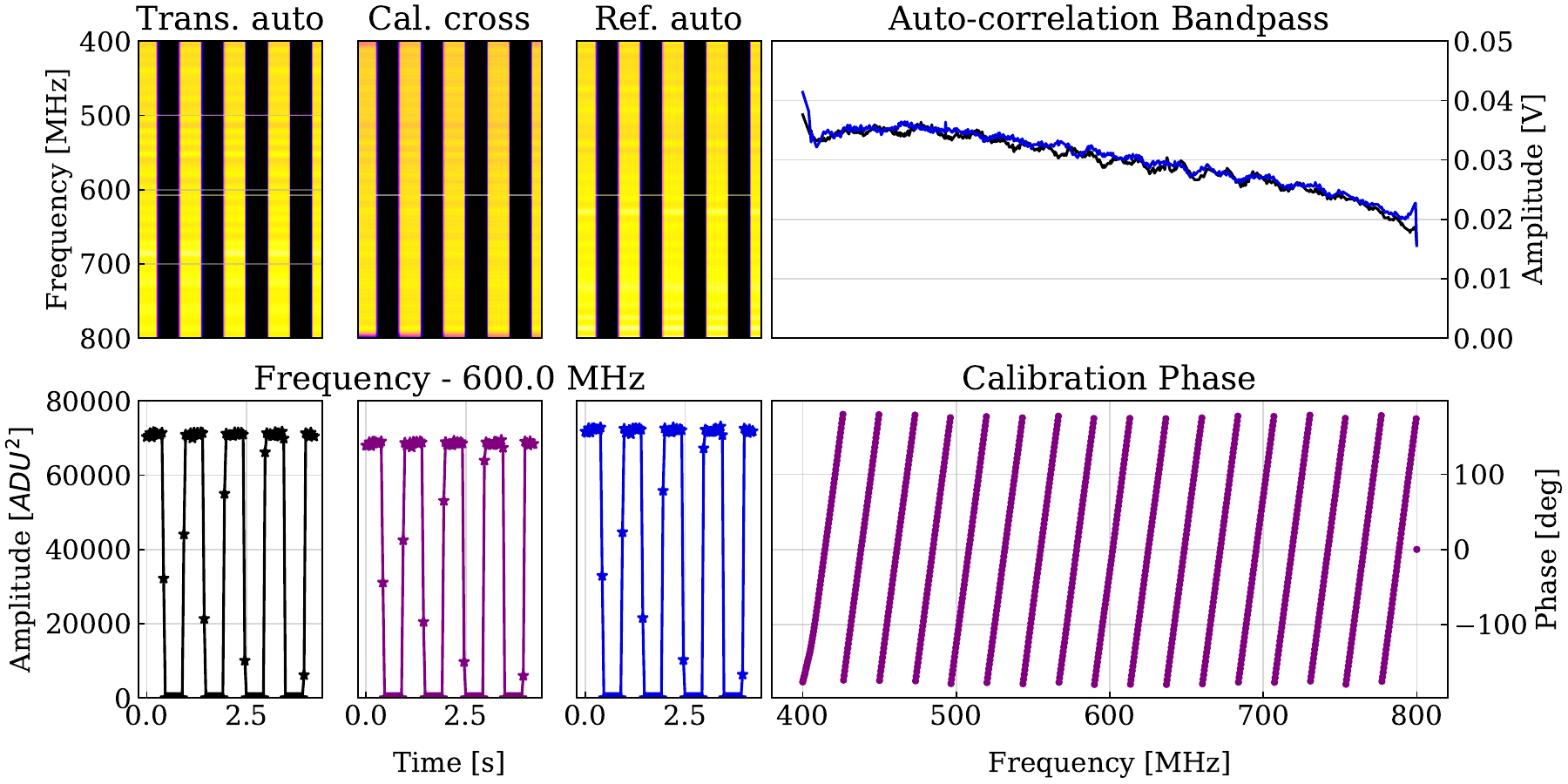}
\caption{\small{\textit{\textbf{(Upper left)} Waterfall representation of the 1\,s periodic pulsed calibration signal over $\sim4$\,s. Each panel shows the same time period for the auto-correlation (left), correlated channel (middle) and reference auto-correlation (right). For left panels, the horizontal axis is time spanning over $\sim4$\,s and for right panels it is frequency spanning 400-800\,MHz. In the correlated channel, the absolute value is shown since the signal is complex. The synchronized pulsing seen in the correlated channel across all frequencies is evidence that the calibrator source is functioning as expected. \textbf{Lower left} This shows the signal as a function of time for a single frequency, with the same time period as above. For the correlated channel, only the absolute value is shown. The amplitude is ADU units proportional to power. The SNR shown is $>1\times10^{5}$. \textbf{Upper right} This shows the bandpass for transmission and reference channels for one \son time sample. The slope is due to frequency-dependent response of the analog reconstruction filter located immediately after the DAC. \textbf{Lower Right} This shows the phase angle of the correlated channel for the same time period across the same bandwidth as other panels. The phase exhibits a linear frequency ramp indicating deterministic inter-channel delay and stable wideband synchronization. This figure indicates excellent SNR in the correlated channel with an amplitude that remains spectrally flat across bandwidth, consistent with a broadband, coherent Gaussian signal and low-jitter clocking.}}}
\label{fig:BenchTest_Results}
\end{figure*} 

\section{Benchtop Validation of Calibration Source}
\label{sec:benchtop_valid}

To characterize the performance of the RFSoC-based deterministic noise generation system, particularly its signal-to-noise ratio (SNR) under realistic synchronization conditions, we carried out a series of controlled bench tests. Each RFSoC4x2-based DAC synthesized pseudo-random Gaussian noise across a 1228.8\,MHz bandwidth (centered at 900\,MHz) and we connected each unit to an ICEboard input channel via a 375\,cm coaxial cable. Each calibration source was provided a 10\,MHz square wave and synchronized PPS trigger by an independent Furuno clock set to self-survey mode. The reference and transmission signals were set to produce a 1\,s period, $50\%$ duty-cycle calibration signal synchronized to the PPS signal to provide alternating calibration ( \son) and a noise ( \soff) states with sufficient integration time for high SNR. The transmitted signal measured approximately -98\,dBm/Hz on the spectrum analyzer, corresponding to 26.81 ADC RMS units per 390\,kHz channel, whereas the off signal measured approximately -143 dBm/Hz, corresponding to 0.345 ADC RMS units per 390\,kHz channel. 
The signal was then analog filtered  to 410-785\,MHz to match the operational range of the ICEBoard. During data acquisition, the ICEBoard was configured in \texttt{corr16} mode~\citep{Bandura2016} --- the received signals from both channels were channelized into 1024 frequency bins between 400-800\,MHz and the inputs were correlated and accumulated into 41.94304\,ms correlator dumps. The resulting auto- and cross-correlation products were written to disk. 


Figure~\ref{fig:BenchTest_Results} shows the results from the bench test. The upper left panel contains a waterfall plot showing a strong signal-to-noise in the correlation channel (Cal. cross) indicating successful signal retrieval. The lower left panel shows signal as a function of time for frequency 600\,MHz for the same period of 4\,s. Because identical copies of the signal are delivered to the correlator in a noise-free environment and nearly identical gain corrections are applied to both channels, the strengths of the auto- and cross-correlations are determined by the same underlying signal power and hence are similar amplitudes. Direct SMA connections minimized system and environmental noise, resulting in an extremely high ($>1\times10^{5}$) cross-correlation SNR. The upper right panel shows the bandpass for both transmission and reference channel during a single \son time sample, demonstrating excellent spectral uniformity across the full signal band. Ripples observed are due to reflections in the cables. Moreover, the observed slope across the band is attributed to the frequency-dependent response of the analog reconstruction filter located immediately after the DAC. While this filter effectively attenuates higher-order spectral components present in the upper Nyquist zones, it does not correct for the intrinsic non-linearity within the first Nyquist zone. Although oversampling could mitigate this distortion, it has been deferred due to the current timing constraints and hardware limitations of the RFSoC4x2 platform. Finally, the phase for that same time across the same bandwidth of 400\,MHz is shown in the lower right panel, showing linear frequency ramp structures, reflecting consistent inter-channel time delay and accurate phase retrieval, consistent with the cable delay. 

The cross-correlation power time series in these bench tests exhibited a high signal-to-noise during the \son phases, validating our RFSoC-based implementation of deterministic broadband Gaussian noise generation and recovery.
With both boards synchronized by independent GPSDO clocks, 
the system consistently achieved robust signal correlation across the targeted frequency band and maintained stable correlations continuously for $\sim$90 minutes, demonstrating the long-term timing and gain stability of the module. This test confirmed the ability of the RFSoC4x2 platform to generate, transmit, and recover wideband pseudo-random signals with high fidelity in both amplitude and phase, suitable for wide-band \tcm cosmology instruments.


\begin{figure*}[ht!]
\centering
\includegraphics[width=\textwidth, height=0.5\textwidth]{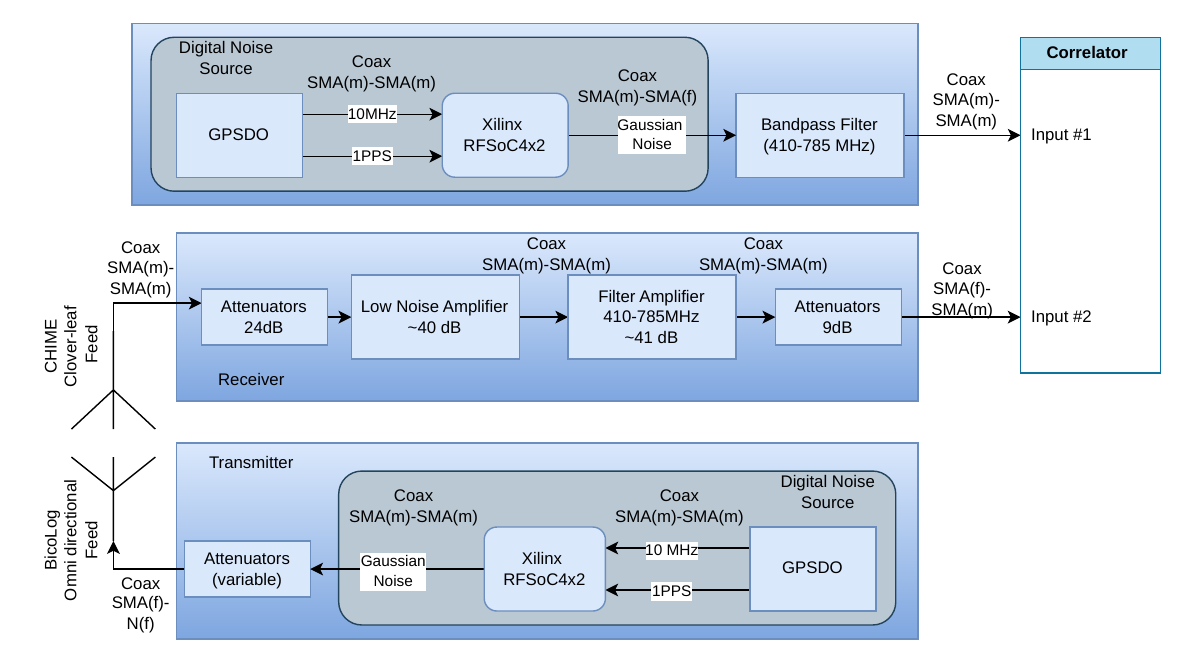}
\caption{\small{\textit{Signal-chain configuration for the anechoic-chamber tests. In this test, a 1.2288 GHz bandwidth pseudo-random Gaussian signal was generated and transmitted from the DAC port of the transmission board, attenuated, and sent via 20\,m coaxial cable to the omni-directional transmitting antenna mounted on a horizontal rod extending from the chamber wall. The transmitter radiates towards a receiving antenna oriented boresight-on and co-polarized. The receiving antenna was a wideband, dual-polarization, cloverleaf-shaped feed, similar to the CHIME antenna, and mounted with a pie-shaped metallic can acting as a ground plane. This antenna was fixed on a motorized rotation stage, which allowed for controlled, incremental rotation. The received signal is first attenuated by 24\,dB to mitigate strong RFI so that amplifier down the chain is not saturated, then amplified by a low-noise amplifier by $\sim40$\,dB. The amplified signal is transported over a 15\,m coaxial running from the chamber to a filter amplifier (FLA)~\citep{Bandura:2014}. The FLA consists of a custom Mini-Circuits bandpass filter (410-785\,MHz) followed by three gain stages. The LNA was DC-powered via the FLA over the same coaxial cable. Finally, the signal is attenuated by 9\,dB to yield an input level of $\sim17$ ADC RMS counts at the correlator front end. Simultaneously, the reference board generated an identical copy of the calibration signal. This reference signal was also filtered through a 410-785\,MHz bandpass filter and fed into a second correlator input.  This setup enabled a measurement of the angular beam response by rotating the receiving antenna.}}}
\label{fig:chamber_schematic}
\end{figure*}

\begin{figure*}[ht!]
\centering
\includegraphics[width=0.8\textwidth]{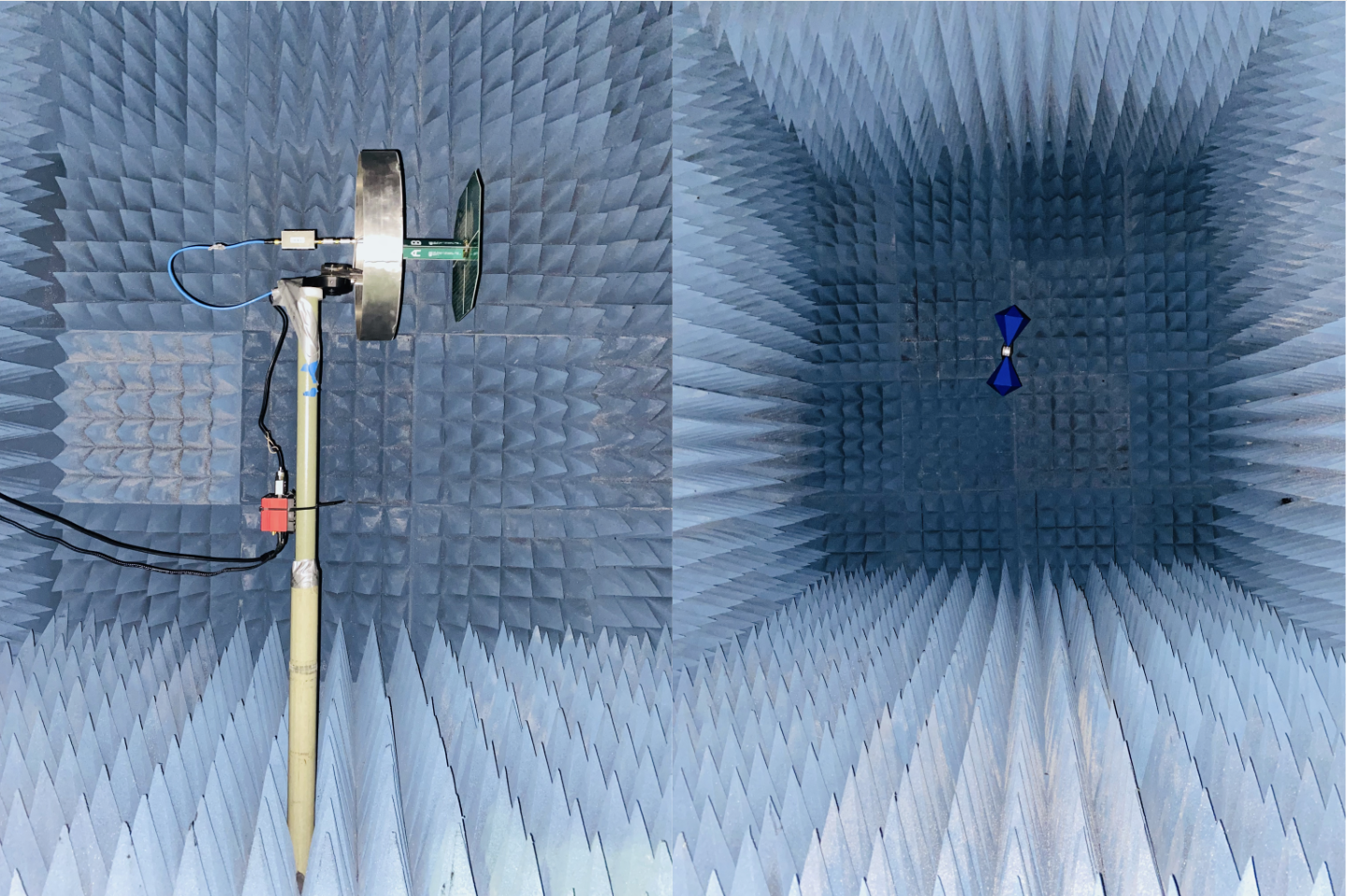}
\caption{\small{\textit{Photograph of the anechoic chamber test configuration showing the Aaronia BicoLOG transmitting antenna (right, 30\,MHz–1\,GHz) radiating towards a CHIME-like cloverleaf feed (left, 400\,MHz–800\,MHz) mounted on a motorized rotary stage. During beam measurements, the cloverleaf feed is rotated while the omni-directional BicoLOG antenna broadcasts the broadband calibration signal. The absorber lining the chamber walls suppresses reflections during measurements.}}}
\label{fig:chamber_setup}
\end{figure*}

\begin{figure*}[ht!]
\centering
\includegraphics[width=\textwidth]{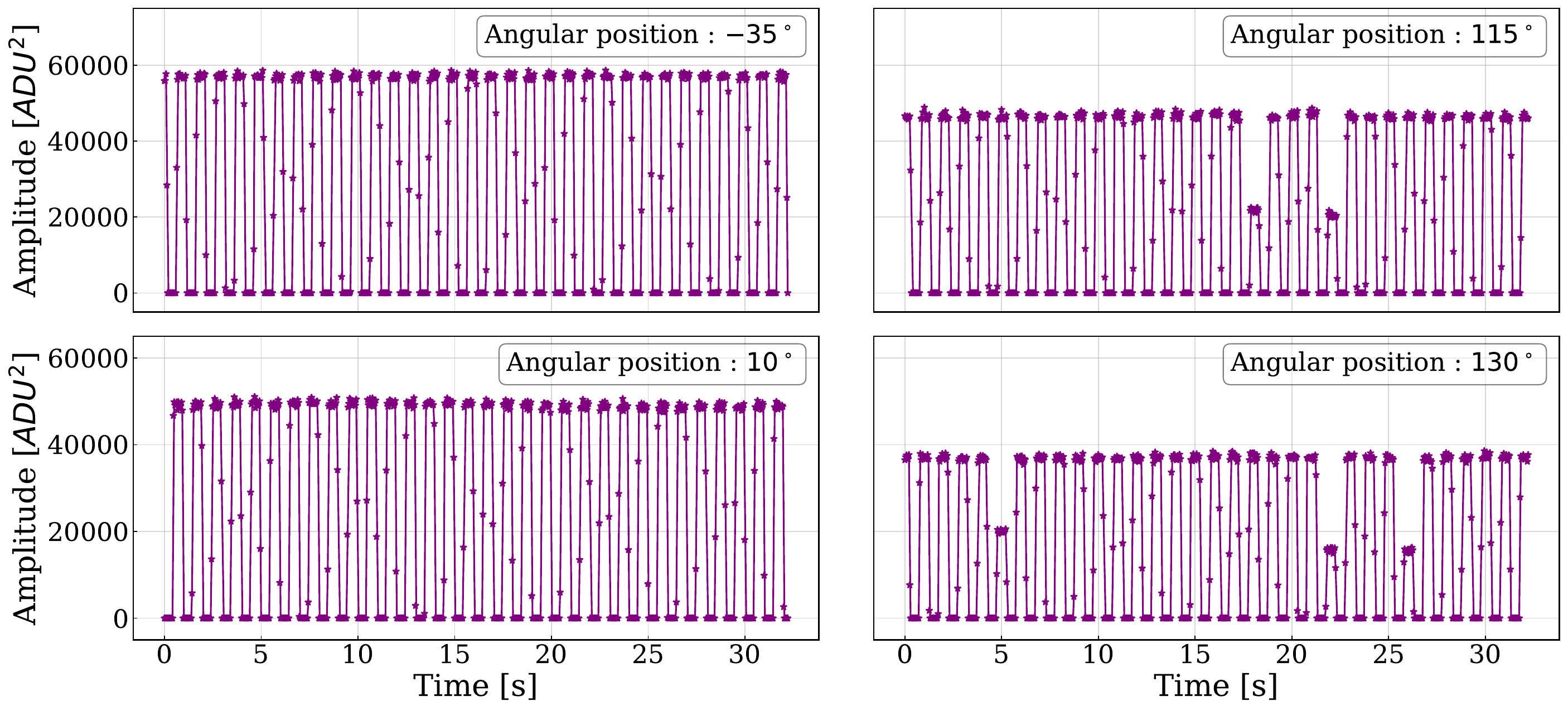}
\caption{\small{\textit{Cross-correlation power as a function of time at four angular positions at 467.97\,MHz. The left column shows time series at $\theta$= $-35^\circ$ (top) and $\theta$= $10^\circ$, where the correlation amplitude remains stable. The right column shows the corresponding time series at $\theta$= $115^\circ$ (top) and $\theta$= $130^\circ$ (bottom), where multi-second dips in the correlation amplitude are clearly evident. These transient features are attributed as instrumental outliers associated with the reference and distribution clock subsystems of the RFSoC platform. They are filtered in the subsequent data analysis as described in Section~\ref{subsec:outliers}.}}}
\label{fig:AnechoicChamber_outliers}
\end{figure*}

\section{Antenna Characterization with Calibration Source in Anechoic Chamber}
\label{sec:chamber_measure}




\subsection{Anechoic Chamber Measurement Setup}

To test the digital source via transmission across free-space instead of through coaxial cables, we chose to characterize the angular response of a radio antenna in an anechoic chamber located at WVU. The experimental setup is summarized schematically in Figure~\ref{fig:chamber_schematic}, with the corresponding implementation inside the anechoic chamber shown in Figure~\ref{fig:chamber_setup}. In this test, 1.2288\,GHz bandwidth pseudo-random Gaussian signal was generated from the DAC port of the transmission board and transmitted from an omni-directional, stationary biconical antenna\footnote{\href{https://aaronia.com/}{Aaronia Bicolog 30100}}. The signal was received in a CHIME-like antenna~\citep{deng14} and sent to an ICEboard correlator input. The second reference board was connected directly to another input of the ICEboard. Each board was synchronized using a separate clock: the clock disciplining the transmission board was set to Navigational mode, while the clock for the reference board was operated in Time-Only mode, as in Section~\ref{sec:benchtop_valid}.  More details of the signal chains are shown Figure~\ref{fig:chamber_schematic} and described in the caption.

We first set the appropriate signal levels by computing a digital gain per frequency in the ICEboard while the receiver antenna views only the ambient temperature absorber as a measurement of system noise. The auto-correlation noise level was set to $\sim13$ ADC counts ($\sigma\sim13$ LSBs), placing the noise floor at roughly 1/10th of the full-scale ADC range.
While the two antennas were pointed at each other (boresight), we set signal on level to be 1.8$\times$ (2.62\,dB above) the system noise floor such that total RMS level rose to $\sim16$ LSBs, corresponding to roughly 1/8th of the ADC dynamic range. This signal-to-noise level was chosen to match operational settings of CHIME-like telescopes. In contrast, the reference board digital gains were determined while we generated a continuous 1228.8\,MHz-wide Gaussian noise signal.  

The CHIME-like antenna was mounted on a rotatable stage and rotated at $5^{\circ}$ increments between $-50^\circ$ and $+135^\circ$ via a Thorlabs precision rotation stage\footnote{\href{https://www.thorlabs.com}{Thorlabs}}. The radio data and corresponding rotation angle were recorded at each position. Data was acquired for a minimum of 30\,s with the correlator integration time set to 41.94304\,ms. As in the bench-top test (Section~\ref{sec:benchtop_valid}), this yields three data products of interest: the auto-correlations of the two inputs (transmission and reference inputs), and the cross-correlation between them. 


\subsection{Instrumental Outliers}
\label{subsec:outliers}

Before presenting main results, we first describe a class of instrumental outliers identified in the sidelobe measurements and the need for their removal from the data. Figure~\ref{fig:AnechoicChamber_outliers} shows the cross-correlation power as a function of time at four angular positions at 467.97MHz. The left column shows time series with stable correlation amplitude throughout, while the right column reveals brief loss-of-coherence “glitches”, seen as a sudden drop in cross-power for an \son period, in a small fraction of integrations for that angle. These events are intermittent and do not repeat at fixed pulse phase or angular position, which suggests an instrumental origin. 

Each signal chain in the present system is locked to an independent GPSDO, with the 10\,MHz reference up-converted individually to 245.76\,MHz and 409.6\,MHz via on-board PLLs in the clock-conditioning stage. These glitches persist even when both noise sources and the correlator are disciplined by a common distributed frequency standard, pointing to on-board PLL and clock-distribution effects as the source. Thus, these glitching instances are identified as instrumental outliers attributable to the reference and distribution subsystem, rather than to the correlator, front-end signal path, or GPSDO clock. If left unfiltered, such outliers propagate directly into the beam map, elevating the uncertainty at the affected angular positions and potentially leading to erroneous conclusions about the beam pattern. In the controlled anechoic-chamber tests, such glitches can be isolated and diagnosed straightforwardly.
These outliers are therefore identified and removed from the subsequent data analysis using a quantitative threshold applied to the cross-correlation ON-period power, flagging integrations that fall below 90\% of the averaged ON-level for each angular position, a value chosen to be consistent with the loss-of-coherence signature shown in Figure~\ref{fig:AnechoicChamber_outliers}. This thresholding could not be applied when the system is flown on a drone, however, because each 1\,s pulse is acquired over a distance of 1-5\,m such that the signal itself can easily change by more than 10\%. The underlying issue was identified and corrected in a later firmware update, implemented after the drone flight measurements presented in Section~\ref{sec:drone_demo}. 


\subsection{Beam Gain and Signal-to-Noise Parameterization}
\label{subsec:G_parameter}

Throughout this paper, lower-case symbols denote voltage-equivalent (amplitude) quantities, while upper-case symbols denote power-equivalent quantities. According to standard radio astronomy convention, power and temperature are used interchangeably --- a system noise power \tsys carries units of Kelvin but represents a physical power spectral density at the input of the receiver. Quantities estimated directly from the measured visibility data are denoted with a hat (e.g. $\hat{g}$), while theoretically derived quantities carry no hat (e.g. $g$, $G$).

The central challenge in drone-based beam calibration is that the signal-to-noise ratio of the measurement varies across the beam --- it is high near boresight and low in the faint sidelobe regions. We therefore characterize the signal-to-noise ratio of the calibration measurement through the beam gain $G(\theta)$, defined as the ratio of the effective antenna temperature of the calibration source to the system temperature, 
\begin{equation} 
G(\theta) = \frac{G_B(\theta) \times T_{\mathrm{\peacc}} }{T_{\mathrm{sys}}} 
\label{eq:gain_def} 
\end{equation}
where $G_{B}(\theta)$ is the beam amplitude of the receiving antenna at angle $\theta$ from boresight, normalized such that $G_{B}(0^\circ)=1$, and \tpeacc is the calibration source power. As defined, $G(\theta)$ is a dimensionless signal-to-noise ratio that encodes both the directional response of the antenna and the strength of the calibration source relative to the system noise floor. We note that in linear units, $G$ carries an implicit power scaling for the auto-correlation (where signal and noise enter as squared voltage amplitudes) whereas for the cross-correlation it carries a voltage scaling and thus denoted by $g$, a distinction that becomes relevant when comparing the two estimators in Appendix~\ref{app:theoretical_framework}.

The estimator of $\hat{G}(\theta)$ follows directly from the auto-correlation voltages. The \soff auto-correlation (measured when the calibration source is \soff) is proportional to the system noise power, which absorbs contributions from both receiver noise and sky noise,
\begin{equation} 
\langle v_\mathrm{T} v_\mathrm{T}^* \rangle_{\mathrm{(OFF)}} \propto T_{\mathrm{sys}} 
\end{equation}
where $v_\mathrm{T}$ is the voltage at the telescope receiver output and angle brackets denote a time average over the correlation,  while the \son auto-correlation receives an additional contribution from the calibration source,
\begin{equation} 
\langle v_\mathrm{T} v_\mathrm{T}^* \rangle_{\mathrm{(ON)}} \propto T_{\mathrm{sys}} + (G_B(\theta) \times T_{\mathrm{\peacc}}) 
\end{equation}
Forming the fractional excess power between the two states and normalizing by the  \soff-state level yields the natural estimator of $\hat{G}(\theta)$ from the calibrated visibility data,
\begin{equation} 
\hat{G}(\theta) = \frac{\langle v_\mathrm{T} v_\mathrm{T}^* \rangle_{\mathrm{(ON)}} - \langle v_\mathrm{T} v_\mathrm{T}^* \rangle_{\mathrm{(OFF)}}}{\langle v_\mathrm{T} v_\mathrm{T}^* \rangle_{\mathrm{(OFF)}}} \equiv G(\theta)
\label{eq:ghat_def} 
\end{equation}
where the equivalence to equation~(\ref{eq:gain_def}) follows from the proportionality relations above. Since numerator and denominator carry identical power units, $\hat{G}$ is a dimensionless ratio.

\subsection{Anechoic Chamber Beam Maps}
Similar to the lab bench tests, the reference and transmission \peacc boards were set to produce a 1\,s period, $50\%$ duty-cycle calibration signal. For each $5^{\circ}$ step, the $\sim30$\,s of data was fit with an ideal square wave using a Pearson-\textit{r} correlation test to identify \son and \soff intervals (`intermediate' intervals when some of the integration time is on and some off are included in the computation and discarded) as in ~\citet{Kuhn_2025}. For a given angle, the mean and standard deviation of the \son points was computed after removing outliers (see Section~\ref{subsec:outliers}). The standard error on the mean of 4 data points was also estimated by dividing the standard deviation by $\sqrt{4}$, chosen to match the $4\times41.94304$\,ms integration period used in the original simulations for estimating signal-to-noise~\citep{Bhopi_2022}, for all three data products. The mean of the \soff points for the auto-correlation transmission channel were also computed to form a measurement of the system noise for each angle.

Although a 1.2288\,GHz wide Gaussian noise band was generated, our measurements were restricted to the 400-800\,MHz range determined by the ICEBoard correlator. During these tests, significant levels of radio frequency interference (RFI) were present, particularly within the 585-600\,MHz, 620-650\,MHz, 670-690\,MHz, and 720-775\,MHz ranges corresponding to the spectrum allocated to LTE/5G and commercial wireless broadband services. In addition, the calibration signal power relative to the system noise was observed to decrease progressively above 585\,MHz due to the system gain profile. As a result, quantitative analysis in this work is restricted to frequencies between 400-585\,MHz. For this specific experiment, no attenuation was required in the transmission chain. 

\begin{figure*}[ht!]
\centering
\includegraphics[width=0.495\textwidth]{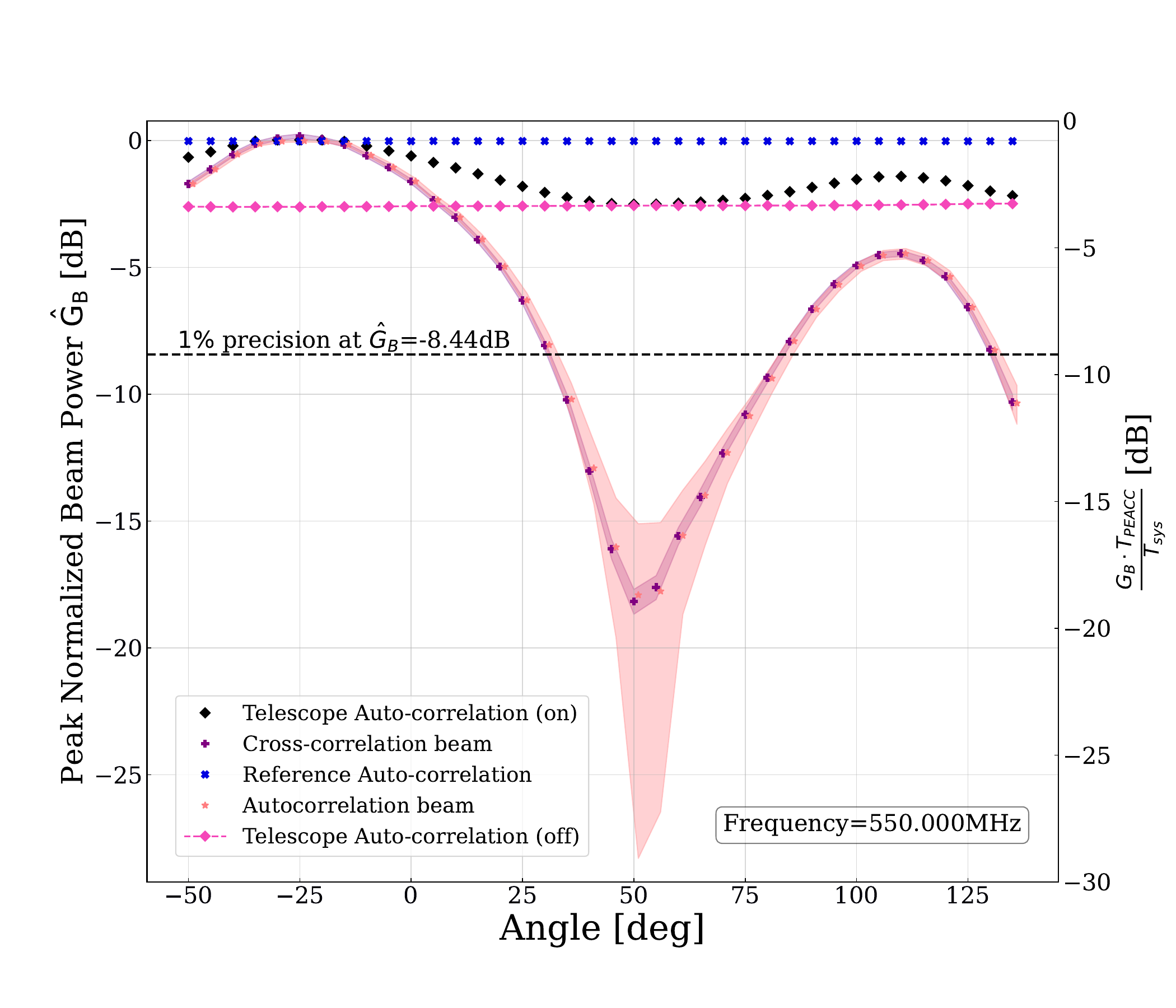}
\includegraphics[width=0.495\textwidth]{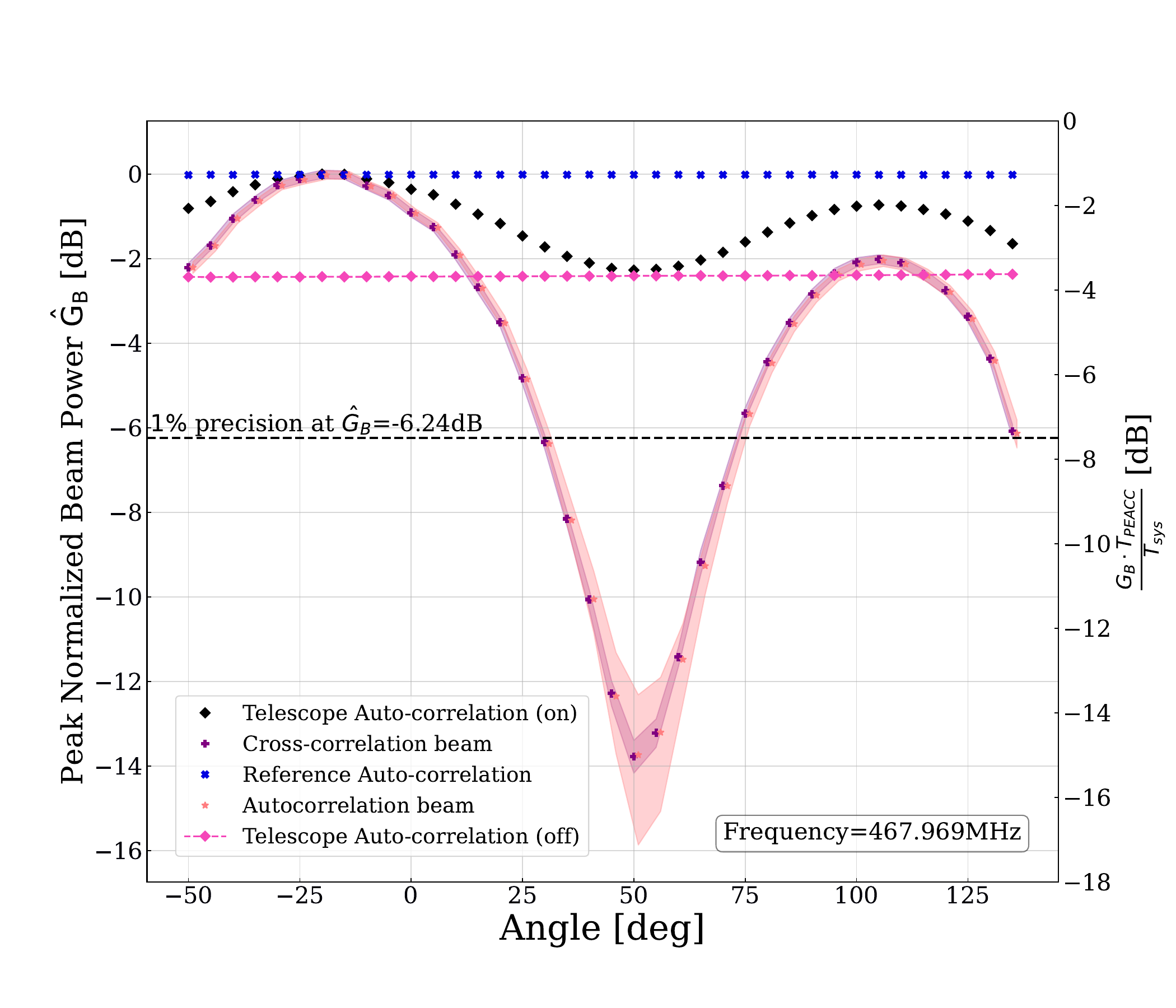}
\caption{\small{\textit{Beam maps for frequency 550\,MHz (a) and 468\,MHz (b), extracted from a 1.2288 GHz–wide Gaussian noise band sampled at 3.6864 GHz with a 900 MHz center frequency (285.6-1514.4 MHz). The left y-axis shows the beam amplitude normalized by the boresight amplitude, while the right y-axis shows the same quantity after noise normalization. $\mathrm{\hat{G}_{auto}}$ and $\mathrm{\hat{g}_{cross}}$ are both plotted, with the former expressed as $\hat{G}_{\mathrm{auto}}\mathrm{[dB]}= 10 \times \mathrm{log}_{10}(\hat{G}_{\mathrm{auto}}\mathrm{[V^{2}]})$ and latter as $\hat{G}_{\mathrm{cross}}\mathrm{[dB]}= 20 \times \mathrm{log}_{10}(\hat{g}_{\mathrm{cross}}\mathrm{[V]})$.  $\hat{G}_{\mathrm{auto}}$ is shifted slightly in angle by +1 degrees to allow for better visualization of the comparison with $\hat{G}_{\mathrm{cross}}$. The measurements demonstrate that, depending on the timing jitter, sidelobes can be measured down to -8.44\,dB and -6.24\,dB, respectively, with $1\%$ precision using the cross-correlation. The correlated results outperform the auto-correlation because they do not contain the system noise.}}}
\label{fig:AnechoicChamber_BeamMaps}
\end{figure*}

The resulting beam maps at two frequencies are shown in Figure~\ref{fig:AnechoicChamber_BeamMaps} and illustrate the angular response derived from both auto- and cross-correlation measurements. Four visibilities are shown: the transmission auto-correlation $\langle v_\mathrm{T} v_\mathrm{T}^{*} \rangle$, which tracks the beam response and reduces to a measurement of $T_\mathrm{sys}$ when the source is  \soff; the reference auto-correlation $\langle v_{\mathrm{ref}} v_{\mathrm{ref}}^{*} \rangle$, which remains essentially constant; and the transmission cross-correlation $\langle v_\mathrm{T} v_{\mathrm{ref}}^{*} \rangle$. From these, the auto-correlation beam amplitude $\hat{G}_{\mathrm{auto}}$ is calculated as:
\begin{equation}
    \hat{G}_{\mathrm{auto}}=\frac{\langle v_{\mathrm{T}} {v_{\mathrm{T}}^*} \rangle_{\mathrm{(ON)}} - \langle v_{\mathrm{T}} {v_{\mathrm{T}}^*} \rangle_{\mathrm{(OFF)}}}{\langle v_{\mathrm{ref}} {v_{\mathrm{ref}}^*} \rangle_{\mathrm{(ON)}} - \langle v_{\mathrm{ref}} {v_{\mathrm{ref}}^*} \rangle_{\mathrm{(OFF)}}} \propto G_{B}(\theta)
\end{equation}
and the cross-correlation beam amplitude $\hat{g}_{\mathrm{cross}}$ is calculated as:
\begin{equation}
    \hat{g}_{\mathrm{cross}}= \frac{\langle v_\mathrm{T} v_{\mathrm{ref}}^{*} \rangle_{\mathrm{ON}}}{\langle v_{\mathrm{ref}} v_{\mathrm{ref}}^{*} \rangle_{\mathrm{ON}}} \propto \sqrt{G_{B}(\theta)}
\end{equation}
Here, $\hat{g}_{\mathrm{cross}}$ is scaled by the reference-channel power which makes it a direct estimator of the voltage beam gain g introduced in Appendix A (Eq. \ref{eq:vton}). Since $g$ is in general complex, we report its magnitude as the beam amplitude here; phase is treated separately in Section \ref{sec:phaseandlims}. Both the $\hat{G}_{\mathrm{auto}}$ and $\hat{g}_{\mathrm{cross}}$  are peak normalized and expressed in decibels as specified in the caption. 
Statistical errors on the mean are computed from the standard deviation (scaled by $1/\sqrt{\mathrm{4}}$ as discussed) for each angle, for both auto- and cross- correlation products. Errors were combined using standard error theory for the $\hat{g}_{\mathrm{auto}}$ error bars. The $2\sigma$ envelopes were constructed by subtracting two standard deviations from the mean at each angle. The 1\% line shown is where the $1\sigma$ statistical error on the mean is 1\% of the beam power.

In Figure~\ref{fig:AnechoicChamber_BeamMaps}, the reference auto-correlation remains essentially constant over all rotation angles, whereas the telescope auto-correlation decreases from boresight toward $\sim-75^\circ$, reaching the system-noise floor, before rising again toward the first sidelobe. The telescope auto-correlation beam powers have large fractional errors particularly at low beam levels, reflecting the intrinsic noise floor of $\sim-2.5$\,dB from the peak, set by the system noise under the present correlator configuration. In contrast, cross-correlation beam powers trace the sidelobes with far smaller uncertainties, consistent with $\mathrm{1/\sqrt{g}}$ noise scaling derived in Appendix~\ref{app:theoretical_framework}, and provide a more robust recovery of the beam amplitude in the low-SNR regime away from the main beam. Under the present system constraints, the measurements show that the recovered beam amplitude remains better than $1\%$ down to a beam amplitude of -8.44\,dB.

\begin{figure*}[ht!]
\centering
\includegraphics[width=\textwidth]{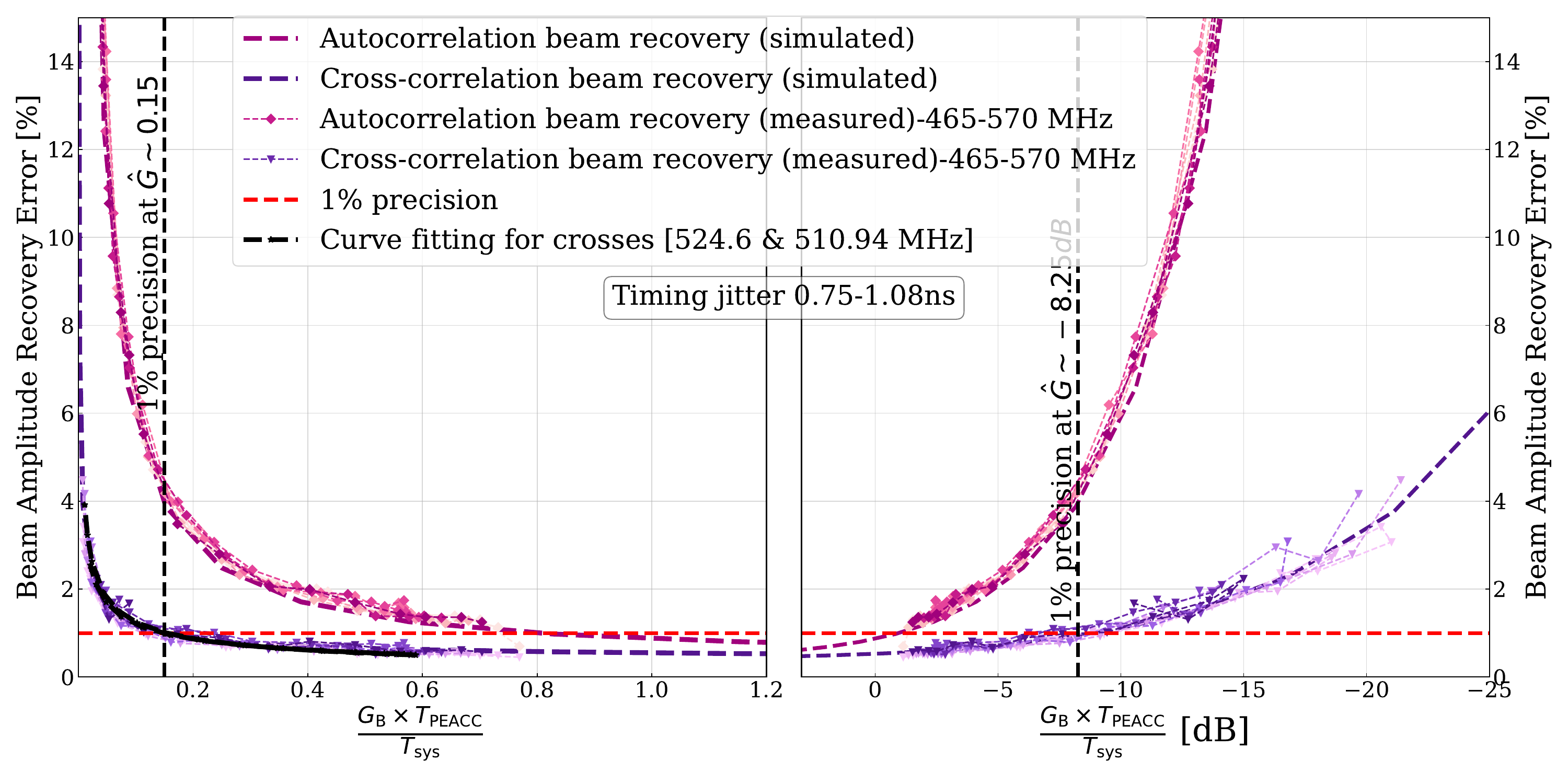}
\caption{\small{\textit{Beam amplitude recovery error as a function of $\frac{G_B \times T_{\mathrm{PEACC}}}{T_{\mathrm{sys}}}$ in linear units (left) and logarithmic units (right) for auto- and cross-correlation estimates for the anechoic chamber data. A horizontal line at 1$\%$ indicates the target precision for beam measurements in the low-SNR regime; In linear space (left), the auto-correlation error falls below the $1\%$ threshold at $\hat{G}$=0.79, whereas the cross-correlation error satisfies this criterion for $\hat{G}$=0.12-0.18, corresponding to timing jitter in the range 0.75-1.08\,ns. In logarithmic space (right), this is equivalent to achieving $1\%$ precision down to -9.21\,dB, demonstrating that sidelobes can be measured to this level under the current system constraints.}}}
\label{fig:AnechoicChamber_BeamAmpRecover}
\end{figure*}

A systematic offset of the CHIME feed beam centroid by approximately $10^\circ$ from the expected boresight was also observed. The origin of this shift is presently unclear; potential contributors are details of the mechanical setup (including any bias in the boresight angle or how we zeroed the reference axis between transmitting and receiving feed), chamber geometry/modeling and residual scattering from instrumentation in the vicinity of the anechoic chamber, but no definitive cause has yet been identified.  


\subsection{Performance Evaluation of the System}
\subsubsection{Achieved System Performance} 

The beam maps are used to infer the effective timing jitter by comparing the measured data to the simulations of ~\citet{Bhopi_2022}. As a brief review of the simulations: two copies of pseudo-random noise signal $s(t)$ were generated, with independent additive noise added to each to form $x_1=g.s(t)+n_1(t)$, $x_2=s(t-\Delta t)+n_2(t)$, where $g$ is dimensionless voltage gain factor representing the signal-to-noise term, varied between 0.005 and 3.2, and $\Delta t$ introduces the simulated timing jitter. The two signals were then auto- and cross-correlated to form $\hat{G}_{auto}=\langle x_1 x_1^* \rangle / |g|^2$, $\hat{g}_{cross}=\langle x_1 x_2^* \rangle / g$, defined such that $\hat{G}=1$ indicates perfect recovery of the input gain, $G \equiv |g|^2$ (or $g$ for the cross-correlation). Since the simulation included both thermal noise and jitter, $\hat{G}$ can be smaller than 1 --- for example, as jitter increases, or as the noise level grows large compared to the signal $T_\mathrm{PEACC}$.

The standard deviation across 4 samples (equivalent to a 164\,ms integration period) was computed. To assess how well $G$ is recovered (or, how close $\hat{G}$ is to 1), standard deviation was divided by $\hat{G}$ (essentially, computing noise over signal). 
A value of 0.01 indicates we achieved a 1\% measurement or better. In the simulations, we varied the jitter to produce heat-maps of noise-over-signal as a function of input gain $G$. In this section, we place our beam map results in these same units to fit for the effective timing jitter by direct comparison to the simulation curves.

As can be computed from the expressions in Section~\ref{subsec:G_parameter}, when $G(0^\circ)=1$, the calibration source contributes power equal to the system noise floor, corresponding to a 3\,dB increase in total received power. The boresight signal level was intentionally set as close as possible to satisfy this condition. This condition is motivated by the simulations of \citet{Bhopi_2022}, in which $G$ is defined identically as the signal-to-noise ratio, setting $G(0^\circ)=1$ therefore places our measurements on the same scale as the simulation curves. This enables a direct fit for effective timing jitter without requiring knowledge of the absolute telescope gain. The measured standard deviation in the data is then compared to the measured $\hat{G}$ at each gain level, and used to fit for timing jitter by comparing to simulations.

The resulting gain recovery estimates $\hat{G}_{\mathrm{auto}}$ and $\hat{G}_{\mathrm{cross}}$, formed as defined in Section~\ref{subsec:G_parameter}, are shown on the right hand axis of Figure~\ref{fig:AnechoicChamber_BeamMaps}. The peak is not quite 0\,dB; this is because the correlator dynamic range limited the \son power to be 1.8$\times$  rather than the target 2$\times$ above the \soff power, placing boresight signal-to-noise marginally below 1.


Figure~\ref{fig:AnechoicChamber_BeamAmpRecover} presents the fractional uncertainty on the recovered beam amplitude ($\frac{\delta G}{G}$) for both the auto- and cross-correlation as a function of the gain estimate $\hat{G}$. Presenting the results in this normalized form not only simplifies the jitter fitting by casting both the data and the simulation curves of \citet{Bhopi_2022} in consistent units, but also makes immediately apparent that the cross-correlation outperforms the auto-correlation at all gain levels; this holds even at large beam amplitudes where the auto-correlation noise would otherwise be expected to be subdominant. Overlaid on the data are simulation curves from \citet{Bhopi_2022}, which predict noise-to-signal as a function $\hat{G}$ for different timing jitter values. 

The percentage error on both the auto- and cross-correlation beam measurements is well described by a power-law scaling with gain $G$, of the form $\delta G = A \cdot G ^{-k} $, with coefficients of determination $R^2 > 0.99$ and $R^2 > 0.91$ respectively across all frequency channels examined. For the auto-correlation, the inverse-variance weighted mean of the fitted exponents across frequency channels is $\bar{k}_{auto}=0.964$, and for the cross-correlation $\bar{k}_{cross}=0.444$. Both values are slightly but systematically below their theoretical predictions of $k=1.0$ and $k=0.5$ respectively, consistent with finite $T_{\mathrm{PEACC}}/T_{\mathrm{sys}}$ corrections that arise when the strict limit $T_{\mathrm{\peacc}} << T_{\mathrm{sys}}$ is not perfectly satisfied. The agreement between the observed power-law scalings and the theoretical predictions confirms that the measurements are in the gain regime where the simulation-based noise scaling holds, supporting the validity of the subsequent jitter extraction. 
Least-squares fits of the power-law form $\delta G = A \cdot G ^{-k} $ to the data are performed independently for each frequency and overlaid on the simulation heat-maps; the effective timing jitter is then extracted by identifying the simulation curve that best matches the fitted exponent and amplitude. The best-fit curves for frequencies 510.94\,MHz and 524.6\,MHz are shown in Figure~\ref{fig:AnechoicChamber_BeamAmpRecover}. Across the full range of frequencies examined, the fits yield effective timing jitters in the range 0.75–1.08\,ns, all comfortably below the 1.7\,ns requirement. We do not quote formal uncertainties on the fitted jitter values, since plausible fitting errors would not bring any value into tension with this threshold nor affect the conclusions drawn from these measurements. 

The fits indicate that timing jitter is not a limiting factor in this regime. As Figure~\ref{fig:AnechoicChamber_BeamAmpRecover} shows, the noise-over-signal curves become insensitive to jitter almost immediately upon the cross-correlation outperforming the auto-correlation, such that further reduction in jitter provides no meaningful improvement. In this regime, improvements in measurement precision are better achieved through increased dynamic range and signal level rather than tighter clocking. The recovered jitter range is furthermore consistent with independent lab measurements of relative jitter as described in Section~\ref{sec:precision_clocking}, providing a cross-validation of both the beam measurement methodology and the clocking characterization. This represents the first validation of free-space coherent calibration of the \peacc system, and demonstrates improved signal recovery at low SNR relative to the auto-correlation measurements. 



\subsubsection{Projected System Measurement Capabilities}
For the tests presented in this paper, we use the ICEboard correlator and operate in a regime where both the signal and the noise are measured in the linear quantization regime where quantization bias in the measured correlations is negligible. As noted above, we set the system noise to occupy roughly 1/10th of the ADC range and the signal-on RMS level to about 1/8th of the range, which allows us to set the signal-to-noise level to $\sim 2$. However, this configuration already saturates the available dynamic range of the ICEBoard correlator: the maximum achievable signal-to-noise ratio was 0.8 instead of 1, with the limit set by the correlator mode of the ICEBoard.

Without this dynamic range constraint, the achieved timing jitter would in principle support $1\%$ measurements at the -20 to -30\,dB sidelobe level, compared to our achieved performance of 1\% measurements at -8\,dB. On the basis of these measurements, reaching -20\,dB with 42\,ms integration time would require a boresight SNR of approximately 12 , while -30\,dB would require a boresight SNR of 120. Equivalently, a correlator with $\sim10$\,dB of usable dynamic range would support $1\%$ measurements at -20\,dB and $\sim3.5$-$4\%$ precision at -30\,dB.

\section{Drone beam measurements of local radio dish}
\label{sec:drone_demo}

\begin{figure*}[ht!]
    \centering
    \includegraphics[width=0.25\textwidth]{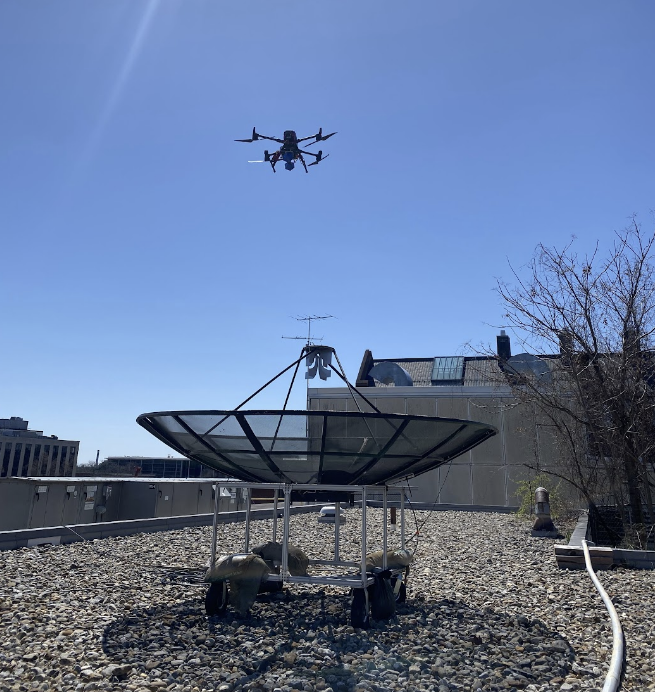}
    \includegraphics[width=0.68\textwidth]{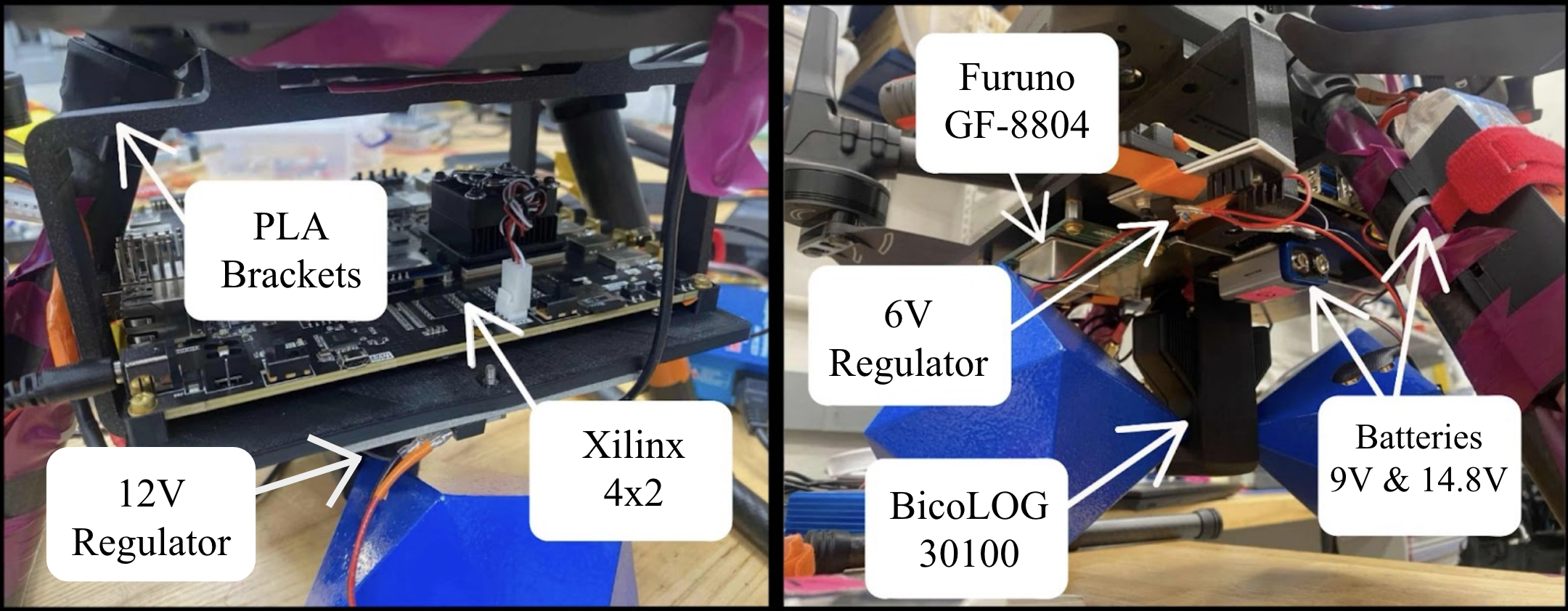}
    \caption{\small{\textit{
    The RTK DJI Matrice 300 RTK drone carrying the full payload hovering over the 3 meter SRT during preliminary centering of the flight trajectory at Yale University (left).
    Top view (center) and bottom view (right) the drone payload attachment with the Xilinx 4x2 board suspended underneath the drone by PLA attachment brackets and platform, along with the BicoLOG 30100 antenna, Furuno GPS clock (GF-8804 model), and batteries. The Furuno requires a 9V battery and the Xilinx board uses a 14.8V battery, which are regulated down to 6V and 12V by linear regulators.
    }}}
    \label{fig:payload_picture}
\end{figure*}

After the anechoic chamber validation tests were completed, we integrated the \peacc onto a drone-carried payload as a means of validating this digital calibrator for radio telescope beam mapping. We performed test flights over a 3\,meter radio dish at Yale University on April 12th, 2025. We made 1D cross-sectional beam maps between 650-700\,MHz, avoiding bright RFI lines.

\begin{figure*}[ht!]
\centering
\includegraphics[width=\textwidth]{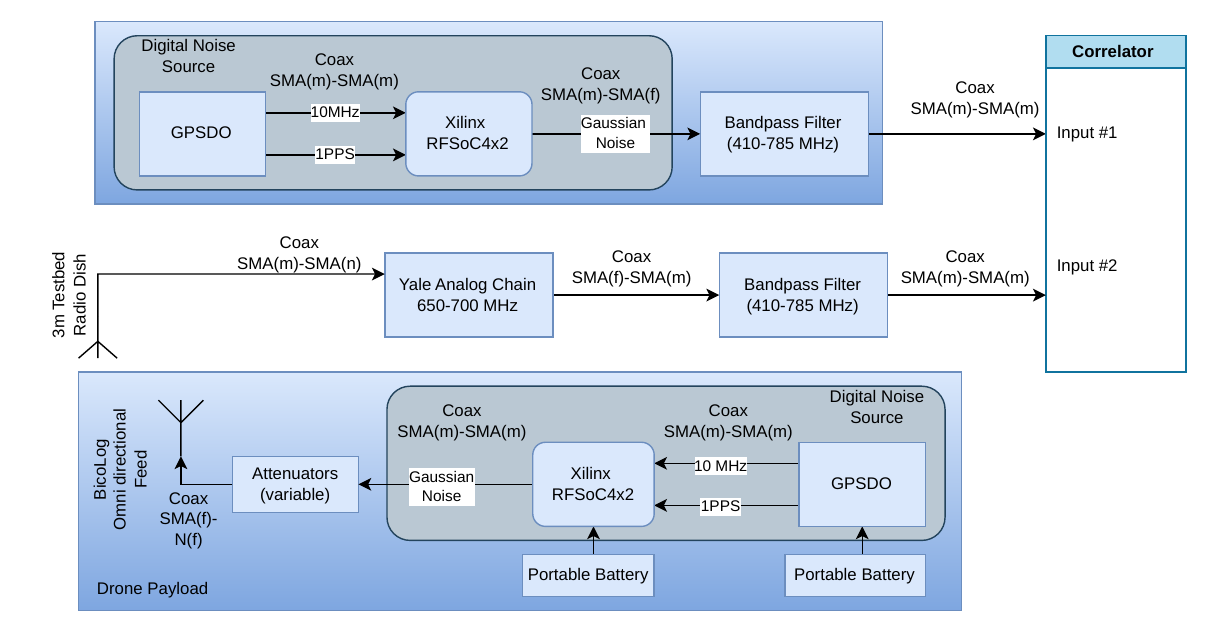}%
\caption{\small{\textit{Signal chain configuration for the drone-based calibration measurements. A 1.2\,GHz-wide Gaussian noise signal is transmitted from the drone using a BicoLOG antenna, with the received calibration power level adjusted via attenuators between the DAC output of the noise source and the transmitting antenna. The BicoLOG illuminates the 3\,m radio dish, which is pointed boresight-on and co-polarized in the North–South direction, and the received signal is carried over a 60 m coaxial cable to the Yale analog backend, comprising a series of amplifiers and SAW filters that define a clean 650–700\,MHz band and a subsequent 400–800\,MHz bandpass stage, yielding an input level of $\sim14$ ADC RMS counts at the correlator front end. An exact replica of the Gaussian noise from the reference source is independently bandpass filtered to 400–800\,MHz and routed directly to a second correlator input.}}}
\label{fig:Yale3m_schematic}
\end{figure*}

The measurement setup was configured similarly to the setup for the anechoic chamber measurements described in Section~\ref{sec:chamber_measure}, as shown by the diagram in Figure~\ref{fig:Yale3m_schematic}. As before, a reference signal was generated by a reference board and received directly by an ICEboard correlator channel. A second, identical signal was independently generated and transmitted by a transmission board mounted on a drone, which flew over a receiver dish that was connected to a second correlator channel. Most board-specific parameters remained the same as the chamber measurements, however, the calibration central frequency carrier was shifted from 900\,MHz down to 200\,MHz. This change was necessary because the 1.2288\,GHz–wide calibration signal produced a spectral image in the second Nyquist zone that overlapped with the drone's control/telemetry band at 2.4\,GHz, while the drone's alternative 5.8\,GHz control band experienced significant interference from nearby construction activities. With a 200\,MHz carrier, the corresponding spectral folding extended only up to $\sim 400$\,MHz and did not contaminate the science band of interest (650–700\,MHz) used during these tests. 

\subsection{3\,m Dish and Receiver Instrument }
These initial drone-based digital calibration tests were performed at Yale’s Wright Laboratory using a 3 meter Small Radio Telescope (SRT) model made in 2004\footnote{\href{https://www.haystack.mit.edu/haystack-memo-series/srt-memos/}{MIT Haystack Small Radio Telescope (SRT)}}. The dish was instrumented with a wide-band prototype feed from CHORD~\citep{2023JAI....1250008M} at the focus. The signal chain was comprised of a long 60\,m LMR-400 coaxial cable from the feed to a surface acoustic wave (“SAW") bandpass filter (650-700\,MHz), followed by a 15\,dB gain amplifier, 3\,dB attenuator, second SAW bandpass filter (650-700\,MHz), low pass filter, 13\,dB gain amplifier, and short SMA cable to the ICEBoard correlator input. This signal chain was carefully designed to operate between 650-700\,MHz to avoid RFI from the densely populated surroundings. The dish was roughly oriented via cell-phone compass such that its two orthogonal polarization axes were oriented due North-South and East-West. For this measurement, only the North-South oriented polarization was acquired in the ICEboard correlator. As in previous sections, the relevant data sets from the ICEboard correlator are two auto-correlation data sets (one for the reference signal and one from the telescope), and their cross-correlation. Also as before, we set an integration time of the correlator to $\sim$ 42\,ms.  
The ICEboard gains are calculated similarly to Section~\ref{sec:chamber_measure} where power from the background (the drone's RTK, sky temperature, system noise, and RFI) is set to roughly 1/10th of the ADC range for the telescope, while the reference board gains were determined using a continuous 1228.8 MHz-wide Gaussian noise signal. We set the power level from the transmission source by flying the source centered over the dish at 50\,m flight altitude, and changed attenuators between the transmission source and transmitting antenna until the signal did not appear to saturate the correlator's dynamic range. To follow a similar analysis as the anechoic chamber measurements, we found an attenuation of 35 dB set the received calibration signal power to be 1.6$\times$ (2.04\,dB) the system noise power while the drone was hovering at boresight.

\begin{deluxetable*}{ccccc}

\tabletypesize{\normalsize}
\setlength{\tabcolsep}{15pt}      
\renewcommand{\arraystretch}{1} 
\tablecaption{Flights presented in this paper, indicating the number of 
passes over the dish in each flight and the heading of the drone per pass. 
\label{tab:flight_dt}}
\tablehead{
\colhead{\textbf{Speed [m/s]}} & \colhead{\textbf{\# Passes}} & 
\colhead{\textbf{Trajectory}} & \colhead{\textbf{Time Offset [s]}} & 
\colhead{\textbf{Position Offset [m]}} 
}
\startdata
0.1 & 1 & Southbound    & $-9.10$ & \nodata \\
1.0 & 1 & Southbound    & $-0.76$ & \nodata \\
2.5 & 2 & South+North   & $-0.76$ & 0.57    \\
5.0 & 2 & South+North   & $~~1.98$ & 0.91    \\
\enddata
\tablecomments{``Southbound'' indicates a southbound pass and 
``South+North'' indicates first a southbound pass then a northbound pass. 
Per-flight timing offset corrections are calculated from manual alignment 
of the measured signal peak to the nominal dish location, i.e.\ the 
drone's closest point of transit over the dish. In the 2.5 and 5\,m/s 
flights, which contain two passes, we present the optimal timing 
corrections that align the latitudes of the signal peaks of the two 
passes, then take the difference from the dish latitude as the position 
offset.}
\end{deluxetable*}

\subsection{Drone and Payload}

The test flights were conducted with a commercial drone, the DJI Matrice 300 with Real Time Kinematic (M300 RTK) GPS\footnote{\href{https://dl.djicdn.com/downloads/matrice-300/20200507/M300_RTK_User_Manual_EN.pdf}{DJI M300 RTK User Manual}}. This platform includes an RTK ground station for drone telemetry. The M300 RTK devices (the drone, controller, and RTK ground station) communicate via a 2.4\,GHz and 5.8\,GHz telemetry link, outside of the chosen \peacc transmission band. Figure \ref{fig:payload_picture} shows the payload attached to the drone, which consists of custom 3D-printed brackets, a 3D-printed attachment plate, a BicoLOG 30100 transmission antenna, the Furuno GPSDO, and 9V and 14.8V rechargeable batteries for the Furuno clock and Xilinx board, respectively. The battery for the Xilinx board is attached to the leg of the drone with a 3D-printed sleeve fixture. The 14.8\,V and 9\,V batteries are regulated to 12\,V and 6\,V, respectively, and attached to an aluminum heat sink. The GPS/GNSS patch antenna for the clock is attached to the front of the drone to maintain line-of-site to sky and minimize interference with the drone's communication link. The transmission antenna is attached below the plate to transmit the signal directly below the drone. The DAC output port of the transmission board is connected to the transmission antenna, and attenuators are placed in this link to adjust the signal power level.

The M300 RTK drone is piloted via a wireless “Smart Controller” which is capable of sending real time flight commands as well as pre-programmed flight missions to the drone. As in~\citet{Kuhn_2025}, we generate automated flight missions with the Universal Ground Control Software (UgCS) package\footnote{\href{https://www.sphengineering.com/flight-planning/ugcs}{SPH Engineering UgCS}}. The drone records flight metrics every 0.1\,s and we convert this to a .csv file via the online Airdata UAV website\footnote{\href{https://airdata.com/}{Airdata UAV Website}}. More than 100 fields are saved, including UTC timestamps, longitude, latitude, altitude, drone orientation (roll, pitch, yaw), and velocity in each direction. With this payload attached, the M300 RTK routinely achieved 20-30 minute flight times at 50\,m altitudes during these test flights before the drone and/or \peacc batteries required recharging.

\subsection{Drone Flights}
All data presented here were acquired within the same hour on a Saturday when construction activities were minimal, generating relatively low RFI. Prior to each flight, we ensured that we were obtaining locked correlations between the two pulsing calibration sources. This was accomplished by hovering the drone in the center of the dish while resetting the on-board clock synthesizers of the reference board until correlated pulsing signatures could be clearly seen in the cross-correlated channel. Once this occurred, we executed the desired flight pattern.

The data presented here is from four pre-programmed flight plans, all at an altitude of 50\,m, well into the far-field of the 3\,m telescope at these frequencies. All flights also had a North-facing drone `yaw' angle (corresponding to transmitting North-South polarization) and flew along one-dimensional North-South transits centered on the nominal dish location, extending out to $\pm$ 15\,m. These are given in Table~\ref{tab:flight_dt}. 


\begin{figure*}[ht!]
\centering
\includegraphics[width=\textwidth]{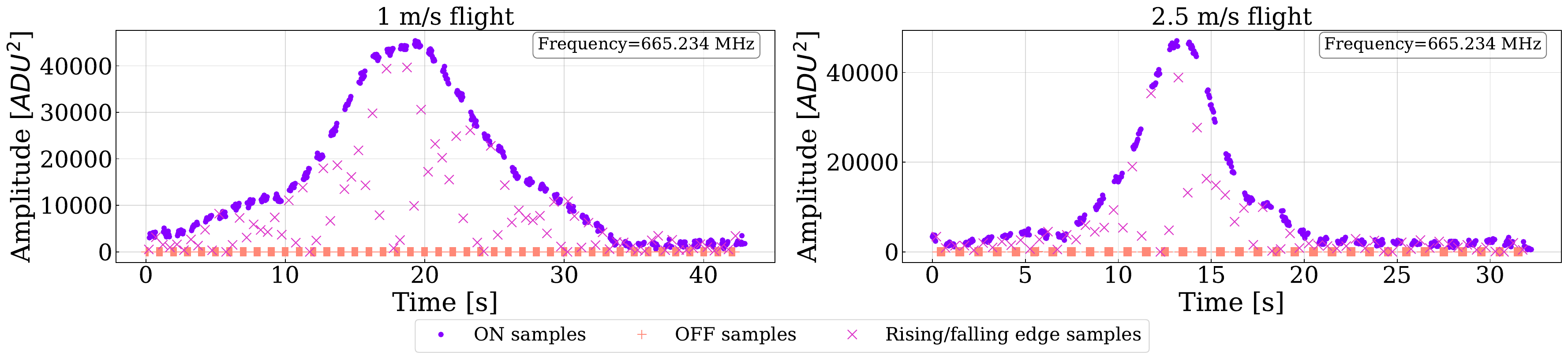}%
\caption{\small{\textit{Single drone transit at 50\,m altitude over the 3\,m dish, recorded cross-correlation spectra (absolute power) at 665.23\,MHz with a drone speed 1\,m/s (left) and 2.5\,m/s (right). Here, medians of the real and imaginary components of the correlated system noise are subtracted from the \son samples of corresponding cross-power spectra to remove constant phase offsets. Shown are median-subtracted \son samples (purple), raw \soff samples (orange), and edge transitions (pink). Median-subtracted \son samples are used for subsequent analysis.}}}
\label{fig:background_subtract_autos}
\end{figure*}


We found the slowest 0.1\,m/s flight data were contaminated by strong local RFI, while the fastest flights do not contain many samples in the beam due to the 50\% duty cycle of the calibration signal, leaving  approximately $50\%$ of the dish beam undersampled. To fill in the beam shape, the 2.5\, m/s and 5\,m/s flights included a second “return" pass over the dish to acquire twice the amount of data. Here, results for the 1\,m/s and 2.5\,m/s are presented after removing data with strong RFI lines (the 0.1\,m/s data) or remain under-sampled (the 5\,m/s flight).  

\subsection{Results and Analysis}

\subsubsection{Drone Data Analysis Procedures}

Following the analysis in Section~\ref{sec:chamber_measure}, we identify \son and \soff data indices in the 50\% duty cycle, PPS-triggered signal. For each \son sample, the background level at that timestamp was estimated from a rolling average of eight nearby \soff samples (four before and four after in time) and subtracted. This procedure was applied to all correlation timestamps and all 1024 spectral channels for the telescope auto-correlation channel. Additionally, the medians of the real and imaginary components of the correlated system noise are subtracted from the \son samples of corresponding cross-power spectra to remove constant phase offsets. The corresponding plots of absolute power are shown in Figure ~\ref{fig:background_subtract_autos} for both the flights.


The next step in the analysis is to synchronize the drone location with the telescope data through their respective timestamps. This is done initially by interpolating the drone data (100\,ms cadence) to the telescope data timestamps (42\,ms cadence). However, because the ICEboard does not use absolute time from the clock, the UTC timestamps produced by the M300 RTK drone could be offset from the timestamps of the ICEBoard data, as found in ~\citet{Kuhn_2025}. This timing offset is constant during the flight, but may vary flight-to-flight. We also assume that there are position offsets to account for (as was found by ~\citet{Kuhn_2025}), including flight-to-flight drone offsets, as well as a dish pointing offset. The nominal dish center longitude and latitude coordinates were determined by hovering the drone at a low altitude directly over the dish, and offsets were then calculated relative to this location. To solve for  timing and spatial offsets, we used the 2.5\,m/s and 5\,m/s flights, which had both forward and reverse passes. Position and time offsets have different signatures in the forward and reverse passes when plotted in drone longitude and latitude coordinates: a position offset looks like an overall shift in the spatial location of the peak amplitude for both passes compared to the nominal dish center. On the other hand, a timing offset looks like an asymmetry in the amplitude peak locations between the forward and reverse flights. We determined an optimal time offset that corrects for the asymmetry in amplitude peak locations of the forward and reverse passes, forcing each peak to a common latitude. We then evaluated how far this latitude is from the nominal dish latitude and took that to be the relative position offset of the drone for this flight. This yielded a unique time and position offset for both the 2.5\,m/s and 5\,m/s flights. It is not possible to distinguish time and position offsets in the single pass flights, thus we only report the time offset for the remaining flights assuming the nominal dish location. The timing offset corrections for all flights are given in Table \ref{tab:flight_dt}. 

We found an overall 0.57\,m and 0.91\,m difference in latitude between the nominal dish location and the 2.5\,m/s and 5\,m/s flights respectively, corresponding to  $\sim$0.7$^{\circ}$ and a 1.0$^{\circ}$ difference at a 50\,m flight height. The data was well-constrained in the latitude direction along the direction of the passes, but not in the longitude direction perpendicular to the passes. These time offsets and updated dish locations are applied to the drone data, which is then re-interpolated to the radio data timestamps. In future tests, these offsets can be derived with higher precision when there is greater spatial coverage per flight.

\subsubsection{Beam Measurement in Amplitude}
\begin{figure*}[ht!]
\centering
\includegraphics[width=0.5\textwidth]{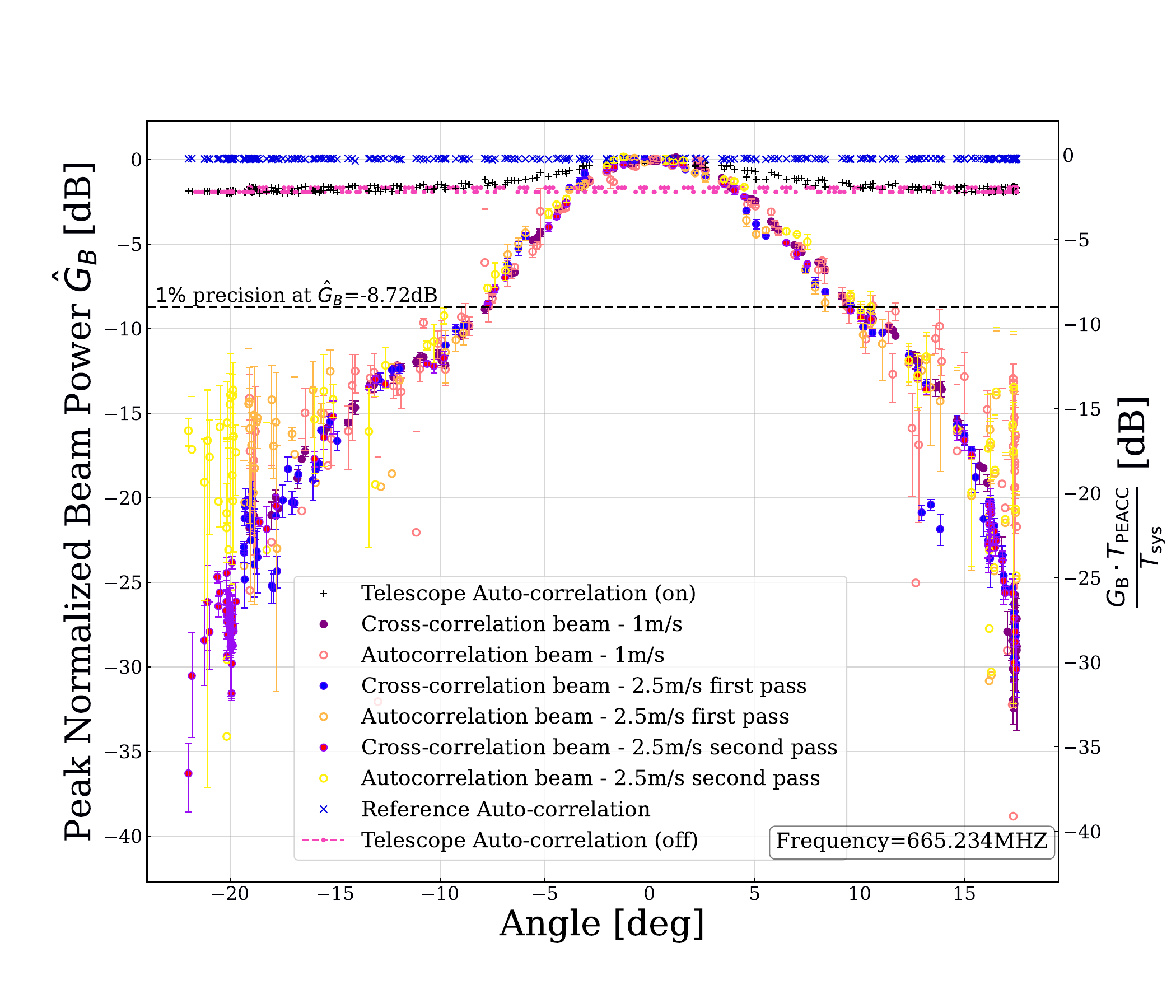}%
\includegraphics[width=0.5\textwidth]{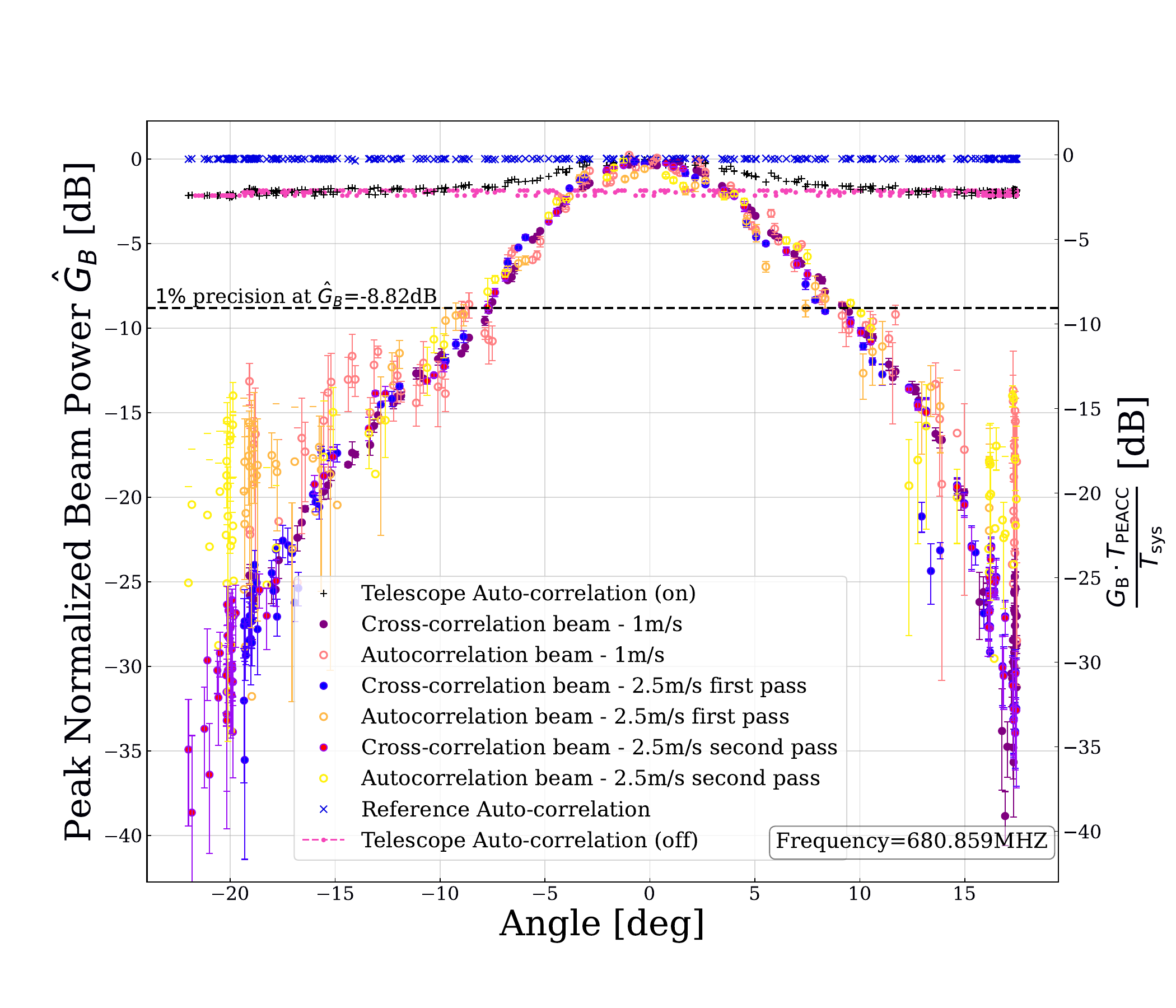}%
\caption{\small{\textit{Beam maps from drone flights at 665.234\,MHz (left) and 680.859\,MHz (right), extracted from a 1.2288\,GHz-wide Gaussian noise band sampled at 3.6864\,GHz with a 200\,MHz center frequency (0–814.4\,MHz). Overlaid are two flights (1\,m/s and 2.5\,m/s) and two drone flight passes (North-bound and South-bound) over the dish for each flight. Data from both the auto-correlation and cross-correlation products are binned into 4 samples and the statistical error is shown as an error bar. The beam profiles are recovered to 1\% precision down to $-8.72$\,dB (665\,MHz) and $-8.82$\,dB (680\,MHz) in the cross-correlated channel, demonstrating the achievable depth as constrained by the system timing jitter and boresight SNR. The autocorrelation data set becomes limited by the system noise floor at small signal levels, while the cross-correlated channel outperforms the autocorrelation at all signal regimes and allows a measurement far lower into the beam shape amplitude. The two flights, and both passes, show consistency within the error bars, indicating good internal agreement on the data set. A more formal statement of agreement is found from the Wilcoxon test, as described in the text.}}} 
\label{fig:Yale3m_BeamMaps}
\end{figure*}

The drone position data is converted to 1D angle using drone height and latitude, collapsing the longitude direction. Resulting beam maps for two frequencies are presented in Figure~\ref{fig:Yale3m_BeamMaps}, which show the angular response of the 3\,m dish as measured in both the telescope auto-correlation and cross-correlation data products. The data are peak normalized, and the error bars are estimated by computing standard deviations over 4 correlator dumps. The auto-correlation and cross-correlation data from the two flights are consistent within error bars. The auto-correlation beam power obtained from the drone-dish channel reaches the system-noise floor at $\sim \pm 10^\circ$, whereas the cross-correlation product does not reach a noise floor and still contains signal down to $\sim$ -30\,dB from peak at $\sim \pm 20^{\circ}$. We find we reach $-8.8$\,dB from peak with 1\% statistical errors and $-30$\,dB with 10\% statistical errors. 

\begin{figure*}[ht!]
\centering
\includegraphics[width=\textwidth, height=0.46\textwidth]{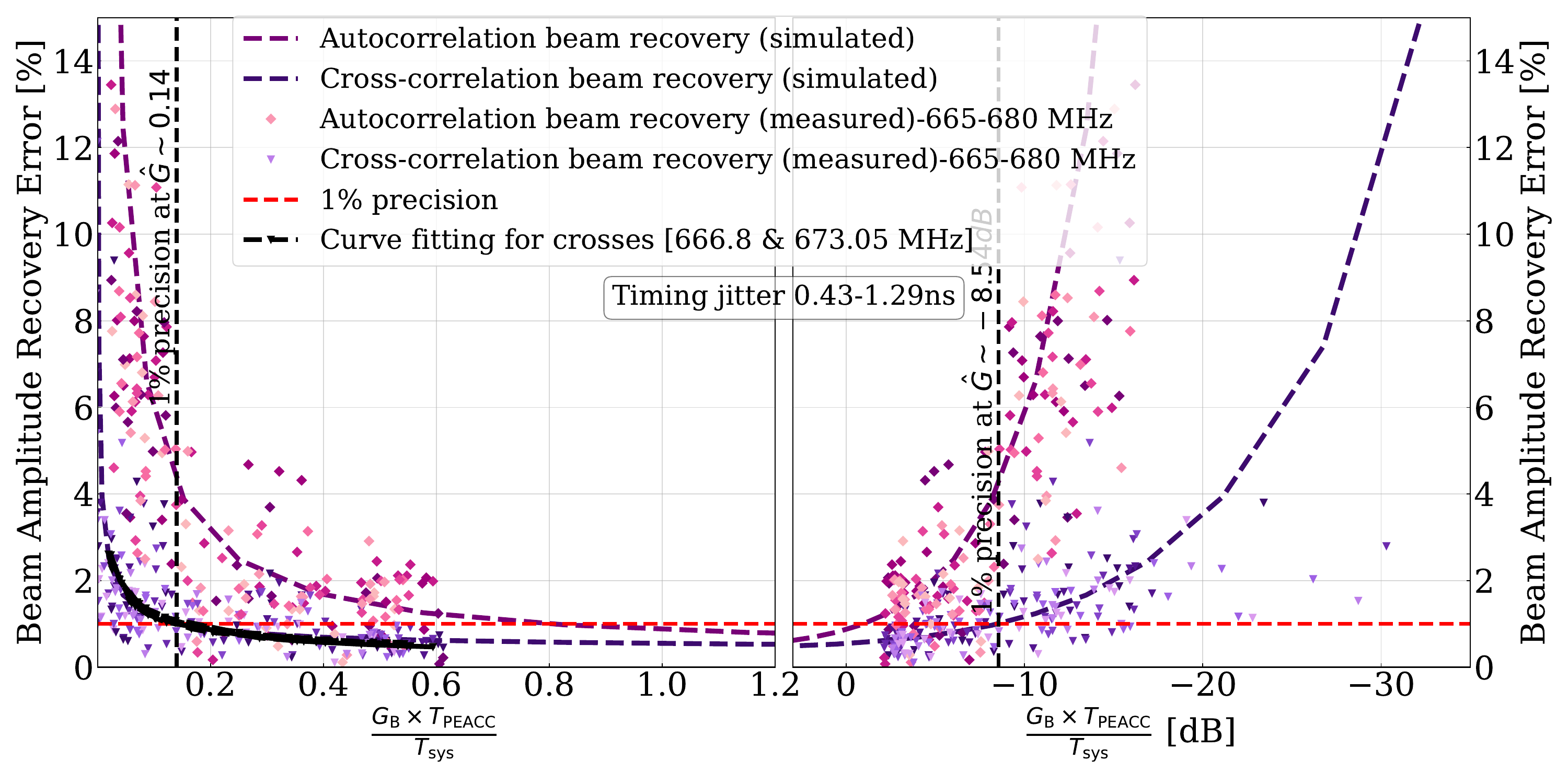}%
\caption{\small{\textit{Beam amplitude recovery error ($\%$) as a function of $\frac{G_{B}\times T_{\mathrm{PEACC}}}{T_{\mathrm{sys}}}$ in linear (left) and logarithmic (right) units for auto- and cross-derived gain estimates (1\,m/s flight). A horizontal red line marks the $1\%$ target precision for beam measurements. In linear space (left), the auto-correlation errors follow the corresponding simulation curve that drop below $1\%$ at $\hat{G}$=0.79, whereas the cross-correlation errors satisfy the $1\%$ criterion at $\hat{G}$=0.14-0.25, corresponding to an effective timing jitter in range of 0.43-1.29\,ns. In logarithmic space (right), this threshold corresponds to $-8.54$ to $-6.02$\,dB, indicating that sidelobes can be measured to this level with $1\%$ precision under the current system constraints.}}}
\label{fig:Yale3m_Results}
\end{figure*}

The beam pattern shown in Figure~\ref{fig:Yale3m_BeamMaps} doesn’t exhibit clear nulls and is wider than expected from a diffraction-limited dish. This could be due to many non-idealities, including whether the feed was at the focus of the dish, uncertainties in the polarization orientation of the dish, and the fact that the feed was designed for a considerably different geometry and f/D ratio. Given the absence of an independent beam model for validation, a direct comparison of the measured beam against theoretical predictions is not currently possible.
Instead, we quantitatively assess system performance by examining the consistency between the auto-derived and cross-derived beam estimates, and repeatability of results across two flights conducted at different transit speeds. This is done using the Wilcoxon test, a non‑parametric, rank‑based test suitable for comparing distributions when the underlying form is unknown and normality cannot be assumed. Applying this test to the three cross‑derived beam estimate distributions from the two flights (single transit for 1\,m/s and two transits for 2.5\,m/s) yields a p‑value of 0.1679, which does not reject the null hypothesis that the distributions are statistically equivalent. In other words, the results are consistent with all three beam measurements being drawn from the same underlying beam pattern. 

The visible consistency between data sets and the results from the Wilcoxon test indicate that we have successfully synchronized two signals, one on a moving platform which we may expect would have larger timing artifacts from the GPS disciplined clocks. In addition, similar to the anechoic chamber measurements, the cross-correlations provide a more robust characterization of the beam response at large angular offsets, particularly in low-SNR regimes. 

\subsubsection{Inferred timing jitter during drone flights}
One major goal of this beam mapping data was to understand if the timing jitter still met requirements even while on a moving platform, where the motion may degrade the timing solutions for the GPSDO clock. To assess this, we repeated the timing jitter inference measurements described in Section~\ref{sec:chamber_measure} for a flight data set. For the 2.5\,m/s flight, each \son and \soff window was further partitioned into groups of four consecutive samples to compute the $\hat{G}$ estimator described in Section~\ref{sec:chamber_measure}. For each four-sample block and for a set of RFI-clean frequency channels, the gain estimators for the auto-correlation and cross-correlation channels were computed following the methodology established in the anechoic-chamber analysis.

The $\hat{G}$ estimator is plotted on the right-hand axis of Figure~\ref{fig:Yale3m_BeamMaps}, with the same error calculations as in Section~\ref{sec:chamber_measure}, where as before, the noise normalization is set by $\hat{G}$  at boresight (angle = 0). Again, we could only achieve 1/8th of the full-scale ADC range, such that the boresight $\hat{G}$  value is -2.2\,dB from peak. To fit the timing jitter as in Section~\ref{sec:chamber_measure}, we recast these data into Figure~\ref{fig:Yale3m_Results}, similar to Figure~\ref{fig:AnechoicChamber_BeamAmpRecover}. Since the $1/g$ and $1/\sqrt{g}$ scalings were validated against chamber measurements in section ~\ref{sec:chamber_measure}, we fix these values here and fit only the amplitude coefficient A as a free parameter; this approach yields more robust fits in the presence of the additional scatter on the data due to the continuously moving drone. Fitting the data to the simulation curves yields an effective timing jitter in the range of 0.43-1.29\,ns over the relevant frequency range, consistent with the \peacc design specification even while on a moving drone platform. The best-fit curves for frequencies 673.05\,MHz and 676.9\,MHz are also plotted in Figure ~\ref{fig:Yale3m_Results}. The measurements in Figure~\ref{fig:Yale3m_Results} confirm that the cross-correlation data provides a better measurement of the beam in all SNR regimes, shown by the fact that the cross-correlation data products lie below the auto-correlation products, indicating lower statistical noise as a function of beam amplitude. As discussed in Section ~\ref{sec:chamber_measure}, we again do not report uncertainties on the fitted jitter values, since the relevant requirement is simply that the effective timing jitter remain below 1.7\,ns. 
\subsubsection{Phase Measurements and Limitations}
\label{sec:phaseandlims}

\begin{figure*}[ht!]
\centering
\includegraphics[width=0.95\textwidth]{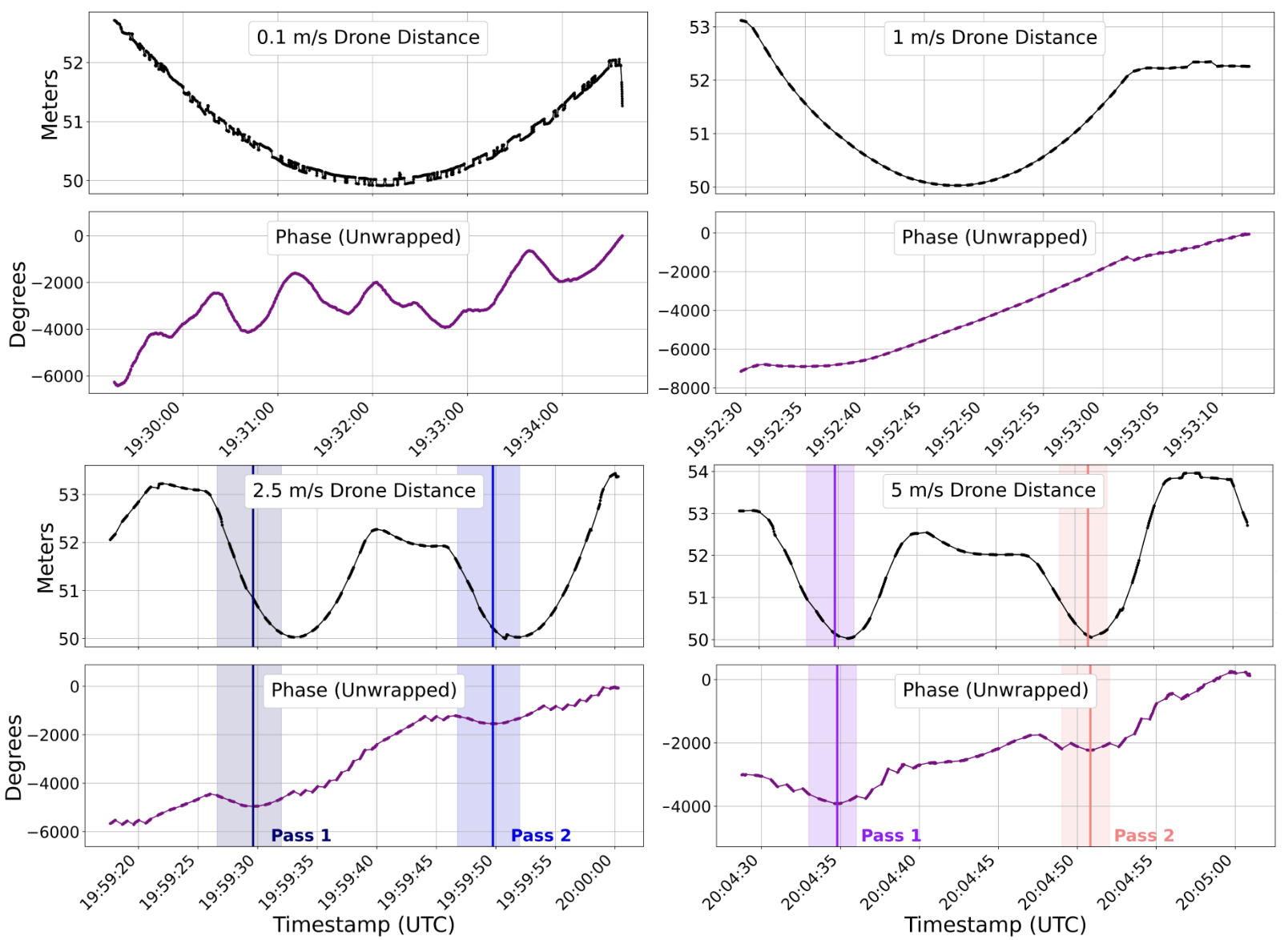}%
\caption{\small{\textit{Distance between drone location and dish center with corresponding measured phase at 663.281\,MHz for each of the 4 flight speeds, plotted against UTC timestamps. We observe that geometric phase recovery seems to improve with faster flight speeds, i.e. on shorter timescales. We see noticeable drifting excursions in the 0.1\,m/s phase which do not appear in the geometry of the drone flight. Similarly, in the 1\,m/s flight we see a steady increasing behavior in the phase data even as the drone moves further away from the dish center. Thus for slow flights, these phase patterns do not reflect the drone's trajectory. The 2.5\,m/s and 5\,m/s flights consist of two passes over the dish, one southbound pass, then the drone stops and changes direction, performing a second northbound pass. In the unwrapped phase, both flights exhibit an overall positive slope unrelated to the drone's trajectory, similar to the 1\,m/s flight. In these two fastest, dual-pass flights, we are able to observe local minima in the phase data (marked by shaded regions and vertical lines). We note that these phase minima all occur prior to the local minima of the corresponding drone distance data. The highlighted regions around these minima reflect the data displayed in Figure \ref{fig:phaseoveramp}.}}} 
\label{fig:phase_v_distance}
\end{figure*}

\begin{figure*}[ht!]
\centering
\includegraphics[width=0.7\textwidth]{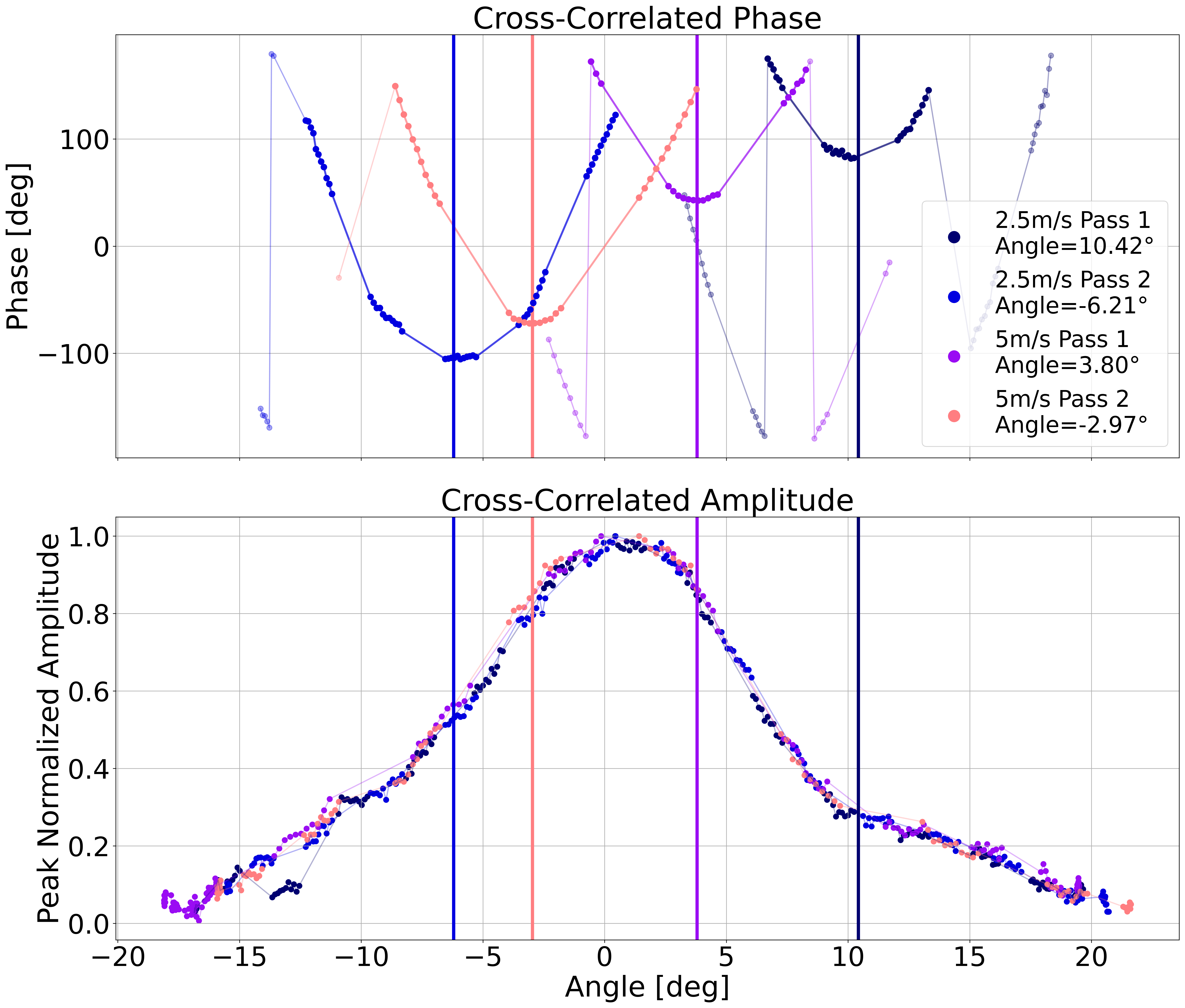}%
\caption{\small{\textit{Cross-correlated complex phase and amplitude at 663.281MHz from the 2.5m/s and 5m/s northbound and southbound drone transits, plotted as a function of drone angle theta relative to the nominal dish center (positive angles north, negative angles south). The amplitude of each transit has been normalized, and thus we can clearly observe that the phase minima are all offset by different amounts relative to the amplitude peak, where we would expect them to fall. In particular, we note that the minima of 2.5 m/s flights fall further from the peak relative to the 5 m/s flights. We also note that the 1st (southbound) passes of each flight fall to the north of peak and the return (northbound) passes fall south of the peak. This implies that phase minima always fall before the amplitude peaks by varying amounts. 
}}}
\label{fig:phaseoveramp}
\end{figure*}

$$
$$
In addition to amplitude, the cross-correlated channel also measures phase ($\Im(v_{\mathrm{T}}v^{*}_{\mathrm{ref}})/\Re(v_{\mathrm{T}}v^{*}_{\mathrm{ref}})$). This phase has two possible contributions: an intrinsic phase associated with the antenna beam response itself, and an instrumental phase arising from the time delay $\Delta t$ between the two inputs (telescope and reference). We write this as:
$$
v_{T}v^{*}_{\mathrm{ref}} = V^{\mathrm{beam}} e^{i \phi}
$$
where $\phi = 2 \pi \nu \Delta t$ captures the instrumental delay-induced phase. $V^{\mathrm{beam}}$ is, in general, complex, encoding both the beam's amplitude and its own intrinsic phase. In the analysis that follows, we find that the beam's intrinsic phase is negligible relative to the instrumental phase, such that the phase results are entirely dominated by the instrumental phase.
We compute the phase as a function of frequency when the drone is at approximately zenith across the different flights and recover a time delay of 0.4\,$\mu$s, consistent with the 66\,m cables (0.85c speed in cables) and 50\,m drone height. Ideally, phase would measure a combination of true beam phase and the phase resulting from the changing time delay between the calibration signal and the drone signal. The time delay is due to both constant path length differences (eg. cables) and changing geometrical offsets as the drone moves across the dish. As a result, we would expect the time delay to track the drone's total distance from the dish, coming to a minimum near the dish's nominal position. 

Measured phase and drone trajectory as a function of time are shown Figure~\ref{fig:phase_v_distance} for all four flights. The trajectory is given in total distance ($r = \sqrt{\Delta x^{2} + \Delta y^{2} + h^{2}}$) from the nominal center of the dish, and the phase has been unwrapped across $\pm \pi$. We find that the slower flights exhibit fluctuating phase (0.1\,m/s) or drifting behavior (1\,m/s flight), inconsistent with the drone's trajectory.  In contrast, in the 2.5 and 5\,m/s flights we observe the phase has a clear minima during the period when the drone was near zenith. Outside of this region, the phase does not clearly follow a geometrical dependence. In the two fastest flights, although the minima in phase exhibit the expected parabolic behavior, they do not quite align in time with the drone distance minima. A simple Gaussian fit was used to determine the phase minima, marked by vertical lines, clearly showing that the phase minima precedes the drone's closest point of transit for all passes. Although only one frequency is represented (663.281\,MHz), we find the phase minima locations are independent of frequency.

For the slow flights, we believe the failure to recover sensible geometric phase indicates that the two reference clocks are not stable on the transit time scales. In Section~\ref{sec:precision_clocking} we observed significant relative time drifts between the two clocks, which were more significant on longer timescales. For reference, we would expect one full $2\pi$ phase wrap just from time drift alone if the time drift was of order $1/\nu$, or about 1.5\,ns for 660\,MHz data. As a result, it is likely that the faster flights (lasting $\lesssim10$ seconds compared to the $\sim$25\,s for slower flights) `outran' the phase drifting from the clocks to yield a phase shape closer to expectations. 

To investigate whether the phase minima had a spatial dependence, we overplot normalized amplitude and phase from the correlated channel against beam 1D angle, derived from the time- and position-corrected drone data. The results are shown in Figure ~\ref{fig:phaseoveramp}. The amplitude data sets have been corrected for timing and dish location, such that the amplitude peaks should align with drone distance, by construction. We find the phase minima are not at consistent locations in the beam, across both passes and both flights. As a result we conclude the phase minima do not correspond to a simple spatial offset, for example from dish pointing. There also seems to be dependence on drone speed: the 5\,m/s flight minima fall closer to the beam peak and are more symmetric around the beam peak than the 2.5\,m/s phase minima. 

The comparison of phase minima to drone location and beam amplitude peaks allow us to rule out some hypotheses: 
\begin{itemize}
\item Because the minima are not in consistent locations, these offsets cannot be due to dish pointing, dish beam artifacts (coma, astigmatism, sidelobes, cross-polarization), or reflections from the environment.
\item Because the drone transmitting antenna orientation was held fixed during flights, these offsets also cannot be due to the transmitting beam shape from the drone. 
\item Changing time delays could slightly decorrelate the signal, leading to a difference between phase minima and amplitude maxima ~\citep{CHIMEholography}, however we would expect this effect to be symmetric in our measurements.   
\end{itemize}

The known and significant drifts in phase found in Section \ref{sec:precision_clocking} are a possible explanation for these differences between amplitude maxima and phase minima by shifting the overall minimum time delay between two inputs. We investigated whether drone `swings' during flights, where the drone tends to tilt towards the direction of travel, could explain this. Although the drone data for tilt angles ruled this out, we were not convinced of the overall tilt offsets (in particular, one changed sign in a way we did not believe), and so this remains a possibility, although we would need to tilt less during the faster 5\,m/s flight compared to the slower 2.5\,m/s flight to match the amplitude and phase data, which is unlikely but possible in windy conditions.  

The phase measurements from these flights indicate we are limited by the stability of the two reference clocks to continue measuring and understanding beam phase. Without improved timing stability, phase can be measured using a reference antenna and could be compared to phase extracted using \peacc. To recover beam phase from this system alone for time scales close to full beam maps with a drone, it will be important to improve relative timing stability.


\section{Future Improvements}
\label{sec:discussion}

We identified a variety of non-idealities and limitations over the course of the measurements. In particular, we found the correlator dynamic range was a significant limitation in our SNR, that the pulsing cadence of 1PPS, 50\% duty cycle severely undersampled the beam when flying at speeds most useful for mapping radio dishes, and finally that the GPSDO timing reference had low jitter on the 1-second reset periods, but drifted significantly on timescales much longer than that, precluding using it as an absolute measurement of phase. Taken in turn: 

\textbf{Correlator dynamic range:} We already discussed in Section~\ref{sec:chamber_measure} the projected measurement performance in the absence of correlator dynamic range limitations. This work therefore highlights two distinct aspects of system performance. First, under the present hardware constraints, the system achieves $1\%$ beam-amplitude precision down to -9\,dB; in this sense, the analysis primarily quantifies the attainable precision as a function of system noise. Second, by considering operation unconstrained by correlator dynamic range, the results indicate that the same architecture could, in principle, support $1\%$ measurements at the -20 to -30 dB sidelobe level, thereby also characterizing the measurement capability as a function of boresight signal level (i.e. dynamic range of the correlator in system chain), a key performance metric for beam-mapping systems. There are a few mechanisms for evading the SNR limits from the correlator, even if using the same ICEboard correlator again. The first is to note that the requirement of a good measurement of noise is not strictly necessary for the beam measurements, this was used only to estimate timing jitter. As a result, we could take advantage of the full dynamic range of the correlator, instead of 1/8th of its range, as was done for these measurements, by setting the gain to the ambient noise as low as feasible. We could also take advantage of dynamically setting the signal levels from the transmission board to increase signal strength for sidelobe measurements. 

\textbf{Undersampling}: If only the 2.5\,m/s or 5\,m/s flights are considered, gaps appear between sampled angles due to the previously mentioned undersampling issue at high flight velocities. The observed sampling pattern suggests two practical strategies: (A) a 1\,s pulse period with slow scanning for dense angular sampling during on-the-fly measurements, or (B) shorter pulses with faster scanning to capture intermediate angles between successive \son windows. These considerations motivate a revised calibration approach in which the digital noise source transmits continuously, with its deterministic sequence reset once per second to maintain temporal correlation. 
This could allow us to run the source continuously, allowing a recovery of the beam shape without any gaps . The proposed next-generation architecture replaces the 1\,s pulsed operation with a continuously transmitted, broadband Gaussian noise signal, while retaining a 1\,s reset of the pseudo-random sequence generator. This continuous-noise mode would allow the drone to traverse the array at higher speeds without loss of correlation, enabling high-fidelity and denser mapping of the synthesized beams of the array elements at the desired spectral resolution and thereby supporting characterization of the telescope as a single, coherently calibrated instrument.

\textbf{Phase}: Ideally the on-board clock has small jitter and low drift, such that it can be used as a phase reference during beam measurements. However, we found significant GPS clock drifting behavior on longer timescales which contaminated the phase data acquired during slower drone transits, and as a result these cannot be used to interpret beam phase geometry (see Section \ref{sec:phaseandlims}). The ideal solution is to implement an alternative highly stable clocking system that is immune to drift (potentially one that is GPS-independent). We may also integrate an additional stationary reference antenna as a secondary diagnostic for geometric phase. 

\textbf{Transmitting beam pattern of the drone antenna}: The transmitting antenna's own beam pattern, both amplitude and phase as a function of angle, is a potential systematic we have not yet fully characterized in this work. In this work the drone's transmitting antenna orientation was held fixed during flights, and the angular offsets observed in Section 5.4.4 could not be attributed to its beam shape, so we do not expect it to have biased the phase results presented here. However, a non-uniform transmit pattern would affect both future absolute phase determination using a stationary reference antenna and the implicit assumption of uniform $T_{\mathrm{PEACC}}$ illumination underlying the sensitivity analysis below. We have since characterized the drone's transmitting antenna in an anechoic chamber as part of a later phase of this project and plan to incorporate this into future work presenting field beam measurements with an improved reference-antenna phase system.

\textbf{Beam sensitivity and flight constraints}: 
The precision with which beam features and sidelobes can be recovered in future drone-based campaigns is governed by two coupled observational constraints apart from boresight SNR: the correlator integration time $\Delta t$ and the dwell time of the drone $\Delta t_{\mathrm{dwell}}$ within the angular extent of the beam feature being measured. From the radiometer equation, the effective uncertainty on the recovered sidelobe level scales as $\delta G_{\mathrm{eff}} \propto (\Delta\nu \Delta t \cdot N_{\mathrm{dumps}})^{-1/2} $, where $\Delta\nu$ is the receiver bandwidth and $N_{\mathrm{dumps}}$ is the number of consecutive correlator dumps averaged. In the present work, we adopt $\Delta t \sim$42\,ms and $N_{\mathrm{dumps}}=4$, giving an effective integration time of $\sim$170\,ms. however, $\Delta t$ cannot be increased arbitrarily --- it must not exceed $\Delta t_{\mathrm{max}} \sim \theta_{\mathrm{res}}/\dot{\theta}$, where $\theta_{\mathrm{res}}$ is the angular width of the beam feature and $\dot\theta$ is the angular velocity of the drone, beyond which source motion within a single dump causes time-smearing that systematically attenuates and biases the recovered amplitude. The dwell time $\Delta t_{\mathrm{dwell}}=\theta_{\mathrm{res}}/\dot{\theta}$ enters independently but symmetrically: since $\delta G_{\mathrm{eff}} \propto \Delta t_{\mathrm{dwell}}^{-1/2}$, the integration time required to recover a sidelobe at a given depth grows quadratically as progressively fainter features are probed, requiring correspondingly slower transits. Slow flights therefore maximize dwell time per feature and relax the time-smearing constraint, whereas fast transits degrade sensitivity in narrow sidelobe features while tightening the upper limit on the permissible integration time. Optimizing the transit speed for a given beam geometry and target sidelobe depth is therefore an important consideration in the design of  future beam-mapping campaigns. We note that this derivation assumes $T_\mathrm{PEACC}$ is uniform across the drone's angular extent relative to the receiving antenna; a non-uniform transmitting beam pattern would modulate the effective calibration signal power with position, introducing an additional systematic beyond the time-smearing effects considered here.

\section{Summary and Outlook}
\label{sec:outlook}

In this paper, we presented the design and implementation of a digital, coherent free-space calibration source that is correlated in radio telescope correlators. We described the underlying motivation for this development and provided a detailed overview of the software architecture for a Gaussian-distributed noise generator. We described the implementation of this source on a Xilinx RFSoC4$\times$2 board, including the GPSDO clocks used for timing synchronization. Lab tests confirmed that the timing system met the jitter requirements of~\citet{Bhopi_2022} and that the RFSoC successfully produced a coherent broadband signal across a 1.2288\,GHz bandwidth. 

We first validated the PEACC module through free-space measurements of an antenna in an anechoic chamber. These measurements demonstrated that the cross-correlation consistently outperforms the auto-correlation across all explored SNR regimes, delivering 1\% precision in beam amplitude down to -9\,dB, and 4\% precision down to -22\,dB. Comparing the data to simulations from~\citet{Bhopi_2022}, we estimated the effective timing jitter to lie in the range 0.75-1.08\,ns, comfortably below the 1.7\,ns requirement and confirming that correlated channel satisfies the design criterion for improved performance over auto-correlation. We note that within this regime, timing jitter ceases to be the dominant source of measurement uncertainty, and further improvements in precision are better achieved through increased signal dynamic range.
Our second validation involved deploying the \peacc system on a moving drone for beam mapping of a 3\,m radio dish. Consistent with the anechoic chamber results, the system produced coherent and unbiased signal injection, and demonstrated improved beam measurements in the cross-correlation channel compared to the auto-correlation channel across all explored SNR regimes. The effective timing jitter recovered from the drone flights was similarly found to be within the range 0.43-1.29\,ns, below the 1.7\,ns requirement. This confirms that the system performs robustly on a moving platform.  

While future improvements remain to be made as discussed, in all cases we found that the correlated channel outperformed the autocorrelation channel, demonstrating the viability of using a digital calibration source capable of generating broadband Gaussian noise at $\sim$GHz frequencies. To our knowledge, this work presents the first published demonstration of free-space coherent signal injection synchronized purely by clocks, the first published deployment of such a source on a moving drone platform, and the first published successful beam measurements made with such a source. Given the growing interest in drone-based calibration for \tcm arrays, this paper establishes the feasibility of using a high-fidelity digital calibration source for next-gen \tcm instruments and provides a practical path towards improved beam calibration in future arrays. 
\begin{acknowledgments}
We gratefully acknowledge everyone who contributed --- both directly and indirectly --- to the development of this module. Special thanks to Danny Jacobs, Cynthia Chang, Jason Gallicchio, Emily Kuhn, and the `drone group' for helpful conversations over the years. We would also like to thank undergraduates Ele Donegan and Andrew Tejada-Vega for their work on the signal chains and cabling for the 3\,m radio dish testbed. We thank the Wright Lab staff for their technical support and assistance throughout this work, particularly during test flights of the 3\,m dishes, who went above and beyond the call of duty to help ensure the infrastructure for the dishes was in place in the unconventional research area of the back hallway and staircase of Wright Lab. We also thank the Leitner Observatory for the use of the dish. This material is based upon work supported by the National Science Foundation under Grants No 1751763, 2108338, 2107929, 2511255, and 2511256.
\end{acknowledgments}

\begin{contribution}
K.~Bhopi led the overall design, implementation, and experimental validation of the \peacc system; developed the simulation framework originally presented in \citet{Bhopi_2022}; implemented the RFSoC firmware; conducted the anechoic chamber and drone flight measurements; performed all data analysis; and wrote the manuscript. W.~Tyndall contributed to the simulation framework development and assisted with clock characterization tests. M.~Cole developed the drone payload and assisted with anechoic chamber and drone flight measurements, data analysis, and manuscript preparation. M.~Helfenbein assisted with the drone payload and drone flight measurements. A.~Whitmer assisted with clock 
characterization tests. K.~Bandura and L.~Newburgh supervised the 
project, provided scientific guidance throughout, and edited the 
manuscript. All authors reviewed and approved the final manuscript.
\end{contribution}

\facilities{WVU Anechoic Chamber, Yale Wright Laboratory: 3\,m SRT,
ICEBoard Correlator}

\software{
\texttt{DigitalNoiseSource}, PEACC firmware and analysis code \citep{DigitalNoiseSource_WVURAIL} and drone-based timing and phase analysis \citep{DigitalNoiseSource_WrightLab},
numpy \citep{harris2020array},
scipy \citep{2020SciPy-NMeth}}

\section*{Data Availability}

The data underlying this article are available at Zenodo, at \dataset[10.5281/zenodo.21997893]{https://doi.org/10.5281/zenodo.21997893}, under a Creative Commons Attribution 4.0 license \citep{Bhopi_2026_Zenodo}. The analysis code used to process these data is available via \citet{DigitalNoiseSource_WVURAIL} and \citet{DigitalNoiseSource_WrightLab}.

\bibliography{main}{}
\bibliographystyle{aasjournalv7}

\appendix


\section{Error Scaling with Beam Gain}
\label{app:theoretical_framework}

We derive how the fractional uncertainty on the recovered beam gain scales with $G$ for both the auto- and cross-correlation measurements. We model the signal received at the telescope antenna as 

\begin{equation}
\label{eq:vton}
    v_\mathrm{T(ON)}= g \times t_\mathrm{\peacc} + t_\mathrm{T,noise(ON)}
\end{equation}
where $t_\mathrm{\peacc}$ is the calibration signal with power $T_\mathrm{\peacc} = \langle t_\mathrm{\peacc} t_\mathrm{\peacc}^*\rangle$, $t_\mathrm{T,noise(ON)}$ is the system noise with power $T_\mathrm{sys}= \langle t_\mathrm{T,noise(ON)} t_\mathrm{T,noise(ON)}^*\rangle$ and $g$ is the voltage beam gain at angle $\theta$ from boresight. The reference signal fed directly to the correlator is $v_{\mathrm{ref(ON)}}= t_\mathrm{\peacc}$, carrying power $T_\mathrm{\peacc}$

\subsection*{A1\quad Auto-correlation}
\noindent The auto-correlation of the received signal is 

\begin{equation}
 S = \langle v_\mathrm{T} v_\mathrm{T}^*\rangle_{\mathrm{(ON)}}= |g|^2 T_\mathrm{\peacc} + T_\mathrm{sys} 
\end{equation}
which follows a chi-squared distribution with variance $Var(S) = 2(|g|^2 T_\mathrm{\peacc} + T_\mathrm{sys})^2$. The beam gain is related to S via $G = |g|^2 = (S-T_{\mathrm{sys}})/T_{\mathrm{PEACC}}$, so $S$ and $G$ differ only by known, fixed scale factors. Averaging over $N_c=2\Delta \nu\Delta t$ independent samples~\citep{Thompson2017}, where $\Delta \nu$ is the receiver bandwidth and $\Delta t$ is the correlator integration time, the uncertainty on $S$ is 
\begin{equation}
    \Delta S = \frac{|g|^2 T_\mathrm{\peacc} + T_\mathrm{sys}}{\sqrt{\Delta\nu \Delta t}}
\end{equation}
In the limit $|g|^2 T_\mathrm{\peacc} \ll T_\mathrm{sys}$, which holds for all but bright beam features, this reduces to $\Delta S \approx T_{\mathrm{sys}}/\sqrt{\Delta \nu \Delta t}$. Since the beam gain is recovered as $G=(S-T_{\mathrm{sys}})/T_{\mathrm{PEACC}}$, the fractional uncertainty on the recovered beam gain propagates directly from $\Delta S$ as
\begin{equation}
    \left(\frac{\delta G_{auto}}{G}\right) \approx \frac{T_\mathrm{sys}}{G\sqrt{\Delta\nu \Delta t}} \propto \frac{1}{G},
\label{eq:auto_scaling}
\end{equation}
where the proportionality holds since $T_\mathrm{sys}$, $T_{\mathrm{\peacc}}$, $\Delta \nu$, and $\Delta t$ are all independent of $G$. 
The auto-correlation percentage uncertainty therefore degrades linearly as progressively fainter beam features are probed. 

\subsection*{A2\quad Cross-correlation}
\noindent The cross-correlation between the telescope and reference signals is 
\begin{equation}
 C =\langle v_{\mathrm{T}} v_{\mathrm{ref}}^*\rangle= g \cdot T_{\mathrm{\peacc}}  
\end{equation}
with variance $Var(C)=2|g|^2 T_{\mathrm{\peacc}}^2+ T_{\mathrm{\peacc}}T_{\mathrm{sys}}$. The voltage beam gain is recovered as $g=C/T_{\mathrm{PEACC}}$, directly analogous to the power gain $G=(S-T_{\mathrm{sys}})/T_{\mathrm{PEACC}}$ from the auto-correlation. In the limit $\mathrm{T_{\mathrm{\peacc}} \ll T_{\mathrm{sys}}}$, the second variance term dominates and the uncertainty reduces to 

\begin{equation}
    \Delta C \approx \frac{\sqrt{ T_\mathrm{\peacc} T_\mathrm{sys}}}{\sqrt{2\Delta\nu \Delta t}}
\end{equation}
Since the cross-correlation power is $g \cdot T_{\mathrm{\peacc}}$, the fractional uncertainty on the recovered gain is

\begin{equation}
    \left(\frac{\delta g_\mathrm{cross}}{g}\right) \approx \frac{1}{g} \frac{\sqrt{{T_\mathrm{\peacc}} T_\mathrm{sys}}}{\sqrt{2\Delta\nu \Delta t}} \propto \frac{1}{\sqrt{G}}
\label{eq:cross_scaling}
\end{equation}
where the $\sqrt{G}= g$ in the denominator yields a shallower scaling than the auto-correlation case~(Eq.~\ref{eq:auto_scaling}). 

Equations~(\ref{eq:auto_scaling}) and~(\ref{eq:cross_scaling}) therefore establish that the fractional uncertainty on the recovered beam gain scales as $\delta G/G \propto 1/G$ for the auto-correlation and $\delta g/g \propto 1/\sqrt{G}$ for the cross-correlation. The cross-correlation is therefore comparatively more sensitive for the recovery of faint beam features at all gain levels. These predictions provide the theoretical basis for the power-law fits to the beam measurement data presented in Section~\ref{sec:chamber_measure}.

\end{document}